\documentclass[12pt]{article}
\usepackage[dvips]{graphicx}
\usepackage{amsfonts}
\usepackage{amsmath}
\usepackage{amssymb}
\usepackage[T1]{fontenc}
\usepackage[latin1]{inputenc}
\usepackage{graphicx}
\usepackage{epsfig}
\newtheorem{prop}{Proposition}[section]
\newtheorem{theorem}{Theorem}[section]
\newtheorem{remark}[theorem]{Remark}

\newtheorem{corollary}[theorem]{Corollary}
\newtheorem{lemma}[theorem]{Lemma}
\newtheorem{proposition}[theorem]{Proposition}
\newtheorem{definition}[theorem]{Definition}

\usepackage{color}

\newcommand{\rd}{\textcolor{red}}

\begin{document}

\title{\phantom{.} {\qquad \hfill \textit{\textbf{\small{J.Math.Phys.}
\rd{\textbf{(07-0355R)}}}}}
\vskip 1.5cm
Lower Spectral Branches of a Spin-Boson Model}

\author{ Nicolae Angelescu $^{1}$, Robert A. Minlos $^{2}$, Jean Ruiz$^{3}$ \\and
\\Valentin A. Zagrebnov $^{4}$}

\date{}

\maketitle

\begin{flushleft}
{\small
${}^{1}$ National Institute of Physics and Nuclear Engineering "H. Hulubei", P. O. Box
MG-6, Bucharest, Romania, e-mail: nangel@theory.nipne.ro
\\
${}^{2}$ Institute for Information Transmissions Problems, Bolshoj
Karetny per.19, GSP-4, Moscow 101447, Russia, e-mail: minl@iitp.ru
\\
${}^{3}$ Centre de Physique Théorique$^*$,
%\footnote{UMR CNRS 6207, Universités Aix--Marseille I et II et Sud Toulon--Var, Laboratoire affilié \`a la FRUMAM}
Luminy Case 907, Marseille 13288, Cedex
9, France, e-mail: ruiz@cpt.univ-mrs.fr
\\
${}^{4}$  Universit\'{e} de la M\'editerran\'ee \ and \ Centre de Physique
Th\'{e}orique$^{*}$ - Luminy, Case 907, Marseille 13288, Cedex 9, France, e-mail:
zagrebnov@cpt.univ-mrs.fr}
\end{flushleft}

\renewcommand{\thefootnote}{}
\footnote{${}^{*}$ UMR CNRS 6207, Universités Aix--Marseille I et II et Sud Toulon--Var,
Laboratoire affilié \`a la FRUMAM}
\renewcommand{\thefootnote}{\arabic{footnote}}

\setcounter{footnote}{0}

\vspace{2em}

{\small\noindent\textbf{Abstract} We study the structure of the spectrum of a
two-level quantum system weakly coupled to a boson field (spin-boson model). Our
analysis allows to avoid the cutoff in the number of bosons, if their spectrum is
bounded below by a positive constant.\\ We show that, for small coupling constant, the
lower part of the spectrum of the spin-boson Hamiltonian contains (one or two)
isolated eigenvalues and (respectively, one or two) manifolds of atom $+ \ 1$-boson
states indexed by the boson momentum $q$. The dispersion laws and generalized
eigenfunctions of the latter are calculated.}

\vspace{1em}


\vspace{1em}


%%%%%%%%%%%%%%%%%%%%%%%%%%%%%%%%%%%%%%%%%%%%%%%%%%%%%%%%%%%%%%%%%%%%%%%%%%%%%%%%%%%%%%%%%%%%%%%%%
\newpage
%%%%%%%%%%%%%%%%%%%%%%%%%%%%%%%%%%%%%%%%%%%%%%%%%%%%%%%%%%%%%%%%%%%%%%%%%%%%%%%%%%%%%%%%%%%%%%%%%%
\tableofcontents
%%%%%%%%%%%%%%%%%%%%%%%%%%%%%%%%%%% Section 1 %%%%%%%%%%%%%%%%%%%%%%%%%%%%%%%%%%%%%%%%%%%%%%%%%
%\setcounter{section}{-1}

\section{Introduction and the main result}\label{sec:1}
\setcounter{equation}{0}

The paper is devoted to the study of the low-lying spectrum of the
Hamiltonian of the so-called "spin-boson" model. The Hilbert space
of the model is
\begin{equation*}
{\cal H}=\mathbb{C}^2 \otimes {\cal F}_s ,
\end{equation*}
where ${\cal F}_s$ is the symmetric (boson) Fock space (see
\cite{Berezin}):
\begin{equation*}
{\cal F}_s={\cal F}_{s}\left( L^2({\mathbb R}^d) \right)
=\bigoplus\limits_{n=0}^\infty {\cal F}^{\left( n\right) },
\end{equation*}
with ${\cal F}^{\left( 0\right) }={\mathbb C}$, ${\cal F}^{\left(n\right) }=
\left( L^2({\mathbb R}^d) \right)_{\text{sym}}^{\otimes n}$ ($n\geq 1$)
the symmetric tensor power, endowed with the scalar product
\begin{equation*}
(\Psi,\Phi)_{{\cal F}^{\left( n\right) }}=\left\{
\begin{array}{lc}
  \psi_0 {\bar \varphi_0}, & {\rm if }\; n=0,  \\ & \\
  ({n!})^{-1}\int\limits _{{\mathbb
R}^{dn}}\psi_n (k_1,...,k_n)\overline{\varphi_n (k_1,...,k_n)}dk_1...dk_n, & {\rm if }\; n\geq 1.
\end{array}\right.
\end{equation*}
The formal Hamiltonian of the "spin-boson" model is defined as an algebraic sum
\begin{equation}\label{Ham-sec-quant}
H = H_0 + H_{int} \ ,
\end{equation}
of the Hamiltonian of non-interacting subsystems of two-level spin and of free boson field:
\begin{equation}\label{Ham-sec-quant-free}
H_0  := \varepsilon \, \sigma_3
\otimes \mathbb{I} + \mathbb{I} \otimes \int_{{\mathbb R}^d} \omega
\left( q \right) a^{*}(q) a (q)dq,
\end{equation}
and of the Hamiltonian coupled these subsystems:
\begin{equation}\label{Ham-sec-quant-inter}
H_{int}  := \alpha \,
\sigma_1 \otimes \int_{{\mathbb R}^d} (\lambda \left( q \right)
a^{*}(q) + \overline{\lambda \left( q \right) }a (q)) dq \ .
\end{equation}
Here,
\begin{itemize}
\item $\sigma _3,\sigma _1$, are the Pauli matrices
\begin{equation*}\label{Pauli-matr}
\sigma_3 = \left(
             \begin{array}{cc}
               1 & 0 \\
               0 & -1 \\
             \end{array}
           \right) \ , \ \
\sigma_1 = \left(
             \begin{array}{cc}
               0 & 1 \\
               1 & 0 \\
             \end{array}
           \right);
\end{equation*}
\item $a^*(k),a(k)$ are the boson creation and annihilation operators
\cite{Berezin}: for $\phi \in L_2({\mathbb R}^d)$, $a^*(\phi )=\int
a(q)^*\overline{\phi (q)}dq$, $a(\phi )=\int a(q){\phi (q)}dq$, which act on the
vector
$$\Psi=(\psi_0,\psi_1(k_1),...,\psi_n(k_1,...,k_n),...)\in {\cal F}_s$$
according to the following rules:
\begin{equation*}\label{a*a}
\begin{array}{ccc}
(a^*(\phi )\Psi)_n(k_1,...,k_n) & = &
        \left\{\begin{array}{lc}
  0, & {\rm if }\; n=0 \ ,  \\ & \\
  \sum\limits _{i=1}^{n}\psi_{n-1}(k_1,...,\check{k_i},...
      k_n)\overline{\phi(k_i)}, & {\rm if }\; n\geq 1 \ ,
\end{array}\right.\\ & & \\
(a(\phi)\Psi)_n(k_1,...,k_n) & = & \int \psi_{n+1}(k_1,...,k_n,k)\phi
(k)dk, \ \ \ \ \ \ \ {\rm if } \; n\geq 0 \ .
\end{array}
\end{equation*}
\item the one-boson spectrum $\omega (k)>0$ (the boson dispersion law)
and $\lambda (k)$ (the form-factor) are functions whose properties
will be discussed below;
  \item $\varepsilon
>0$ and $\alpha \geq 0$ are real parameters, whereby we suppose the coupling "constant"
$\alpha \ll 1$.
\end{itemize}

\noindent The {properties} we require for $\omega (k),\lambda (k)$ are the
following:\\
(A1)\label{ass1} $\omega (\cdot ):\mathbb{R}^d\rightarrow (0,\infty )$ is a
  continuously differentiable function, having a \textit{unique non-degenerate} minimum at the
  origin $\omega (0)=:\kappa $, and with $\partial \omega (q)\neq 0$ for $q\neq
  0$. Moreover, $\lim\limits _{q\to \infty }|\partial \omega (q)|/\omega
  (q)=0$;\\
(A2)\label{ass2} $\lambda\left( \cdot \right):\mathbb{R}^d\rightarrow \mathbb{C}$ is a
  continuously differentiable function, dominated by a bounded positive
  square integrable function $h:\mathbb{R}^d\rightarrow (0,1]$, i.e.
  $$|\lambda\left( q \right)|\leq h(q),\,\, |\partial \lambda\left( q \right)|\leq
  Ch(q),$$ for some $C>0$;\\
(A3)\label{ass3} on every level set $\Sigma _x=\omega ^{-1}(x)$ of the function $\omega$, the function $\lambda
  $ is not identically equal to zero. i.e. for all $\kappa \leq x <\infty $, $\lambda |_{\Sigma _x}\neq
  0$.
%%%%%%%%%%%%%%%%%%%%%%%%%%%%%%%%%%%%%%%%%%%%%%%%%%%%%%%%%%%%%%%%%%%%%%%%%%%%%%%%%%%%%%%%%%%%%%%%%%%%%%%%%%%

\noindent The Hamiltonian $H_0$ of the "free" (non-interacting) system has two
simple eigenvalues $e^0_0=-\varepsilon ,\, e^0_1=\varepsilon $; the
corresponding one-dimensional eigenspaces will be denoted ${\cal
H}_{0,0}^{(i=0,1)}\subset {\cal H}$. Besides, there exist two sequences
${\cal H}_{0,n}^{(i)},\; n=1,2,...$ ($i=0,1$) of $H_0$-invariant
subspaces on $n$-boson states. Under assumption (A1), in each ${\cal
H}_{0,n}^{(i)}$, the spectrum $\Sigma_0$ of $H_0$ is \textit{continuous} and fills the
half-infinite intervals $[\lambda^0_{i,n},\infty )$, where $\lambda^0_{i,n}=e^0_i+n\kappa $
($i=0,1;\; n=1,2,...$).

The paper is concerned with the description of the structure of the lower part of the
spectrum of the weakly interacting ($0<\alpha \ll 1$) system. Namely, our \textit{main result}
(see Theorem \ref{main Th}) can be formulated as follows:
\begin{description}
  \item [(i)] Below the continuous spectrum $\Sigma_0$, there exist either one, $e_0<-\varepsilon $,
  or two, $e_0<-\varepsilon <e_1<\varepsilon$, simple eigenvalues of $H$;
  the corresponding one-dimensional eigenspaces will be denoted
  ${\cal H}_0^{(i)}=\{ \mathbb{C}F_0^{(i)}\} ,\, i=0,1$.
  \item [(ii)] In the orthogonal complement
  $[{{\cal H}_0^{(0)}\oplus {\cal H}_0^{(1)}}]^\perp$ there exist (depending on the number of eigenvalues)
  either \textit{one}, or \textit{two} mutually orthogonal, invariant subspaces ${\cal H}_1^{(i=0,1)}$,
  such that the restrictions $H\mid _{{\cal H}_1^{(i=0,1)}}$ are unitarily equivalent to the \textit{operators
  of multiplication} by the functions
  $$ \mathcal{E}_{i}(q):= e_i+\omega (q) \ , \  i= 0, 1 \ \ ,$$
  acting , respectively, in the Hilbert spaces
  $L_2(G_{\eta }^{(i=0,1)})$, where the domains $G_{\eta }^{(i=0,1)}\subset \mathbb{R}^d$
  are defined by
  \begin{equation}\label{int-spect}
  G_{\eta }^{(i)}=\{ q\in \mathbb{R}^d:\, e_{i}+\omega (q)<\lambda^0_{i=0,n=2}-\eta\} \ , \ i=0,1 \ .
  \end{equation}

  Here, $0<\eta :=\eta (\alpha )$, where $\eta =\eta (\alpha )$ is
  small for small $\alpha $. Thereby, the unitaries establishing the equivalence are
  explicitly constructed.
\end{description}

%%%%%%%%%%%%%%%%%%%%%%%%%%%%%%%%%%%%%%%%% Remark %%%%%%%%%%%%%%%%%%%%%%%%%%%%%%%%%%%%%%%%%%%%%%%%%%%%%%%%%%
\begin{remark}\label{R1.1}
In fact, the two points above exhaust ({though this is not explicitly shown in the
paper}) the spectrum of $H$ in the interval $(-\infty ,\lambda ^0_{0,2}-\eta )$,
meaning that the spectrum of $H$ in the orthogonal complement $\left\{ {\cal
H}_0^{(0)}\oplus {\cal H}_0^{(1)}\oplus {\cal H}_1^{(0)}\oplus {\cal
H}_1^{(1)}\right\} ^\perp $ has no point below $\lambda ^0_{0,2}-\eta $.
{We did not concentrate here on the problem of completeness, although we think that it is possible within out
method. Instead, we focused on the explicit study of the discrete part of the spectrum.}
\end{remark}
%%%%%%%%%%%%%%%%%%%%%%%%%%%%%%%%%%%%%%%%%%%%%%%%%%%%%%%%%%%%%%%%%%%%%%%%%%%%%%%%%%%%%%%%%%%%%%%%%%%%%%%%%%

{Let us briefly describe our method, by which the subspaces ${\cal H}_n^{(i)} \; (n=0,1 \ ; \ i=0,1 )$
and the spectrum of the Hamiltonian $H$ within them are constructed. It consists in the following: consider
the equation}
\begin{equation}\label{eigeneq}
(H-z\mathbb{I})F=0,\;F\in {\cal H},\, z\in \mathbb{R},
\end{equation}
which determines the eigenvectors $F$ and eigenvalues $z$. This
equation can be written as an infinite system of equations for the
components of the vector $F$,
\begin{equation}\label{components}
F=\left\{ f_0(\sigma ),f_1(\sigma ,k),...,f_n(\sigma
,k_1,..,k_n),...\right\},\; \sigma =\pm 1, k_i\in \mathbb{R}^d,
\end{equation}
where $f_n$ are symmetric functions of the variables $k_1,..,k_n$.
After eliminating in a special \textit{effective} way all higher components
$f_n,\, n=2,3,...$ from Eq. (\ref{eigeneq}), we are left with an
equation for the vector $F_{\leq 1}=(f_0,f_1)$ of the form:
\begin{equation}\label{effeq}
A(z)F_{\leq 1}-zF_{\leq 1}=0,
\end{equation}
where $\left\{ A(\xi ),\; \xi \in (-\infty ,\lambda ^0_{0,2}-\eta
)\right\}$ is a family of generalized Friedrichs operators (see,
e.g. \cite{AMZ}). For each given $\xi $, the operator $A(\xi )$ has
one (or two) eigenvalues $e_i(\xi), i=0,1$, which can be
calculated, e.g. as the zeros of the corresponding Fredholm
determinant, while its continuous spectrum is found using scattering
theory and coincides with the spectrum of the operators of
multiplication by certain functions $e_i(q;\xi )$ in the space
$L_2(\mathbb{R}^d)$. Finally, the solutions $e_i$ of the equations
$e_i(\xi)=\xi $ define the discrete spectrum of the operator $H$, and the
solutions $\xi^{(i)}(q)$ of the equations $e_i(q;\xi )=\xi $, which
are shown to be equal to $e_i + \omega (q)$, give its continuous
spectrum.

The method sketched above has been applied in authors' papers
\cite{Minl1}, \cite{AMZ}, where a model of a quantum particle
interacting with a massive scalar Bose field was considered and the
lower branch of the spectrum of its Hamiltonian (\textit{polaron}) has been
studied. It has been also used in \cite{Minl2}, for the analogous problem in a
model of a quantum particle interacting with a massive vector Bose
field (similar to the \textit{Pauli-Fierz} model in electrodynamics).

%%%%%%%%%%%%%%%%%%%%%%%%%%%%%%%%%%%%%%% Remark %%%%%%%%%%%%%%%%%%%%%%%%%%%%%%%%%%%%%%%%%%%%%%%%%%%%%%%%%
\begin{remark}\label{R1.2}
It should be noted that our results concerning the continuous
branches of the spectrum of the Hamiltonian \ref{Ham-sec-quant} are already contained
in the paper \textrm{\cite{DG}}, but there all invariant subspaces of $H$ are constructed
using the abstract methods of \textit{scattering theory}, under condition that its eigenvalues and
eigenvectors are \textit{known}.

{The essential difference is that we construct the eigenfunctions for discrete spectrum
of $H$ as well as generalized eigenfunctions for continuous spectrum (lowest one-boson
spectral branches) \textit{explicitly}.}

{Besides, some analogous results are contained in the papers
{\rm{\cite{Fannes}}, \rm{\cite{Arai}}, \rm{\cite{Spohn2}}, \rm{\cite{Minl3}}}, and also in
{\rm{\cite{G}}, \rm{\cite{Minl4}}}, however in the latter the Hamiltonian $H$ with a
"cutoff in the number of bosons" was considered.}
\end{remark}
%%%%%%%%%%%%%%%%%%%%%%%%%%%%%%%%%%%%%%%%%%%%%%%%%%%%%%%%%%%%%%%%%%%%%%%%%%%%%%%%%%%%%%%%%%%%%%%%%%%%%%%%%%%
%%%%%%%%%%%%%%%%%%%%%%%%%%%%%%%%%%%%%%%%%%%%%% Remark %%%%%%%%%%%%%%%%%%%%%%%%%%%%%%%%%%%%%%%%%%%%%%%%%%%%%
%\begin{remark}\label{R1.3} Notice that the definition of the \textit{ground state} as a certain
%vector $\Psi \in {\cal H}$, which is adopted here and in the papers cited above, is not unique.
%There exist another ways of approaching this notion. For instance, one may consider the KMS state
%(on a suitable algebra $\cal A$ of quasi-local observables) associated to Hamiltonian $H$ at a
%certain temperature $T>0$, and takes its limit $T\to 0$. This approach is adopted in {\rm{\cite{Spohn1}}}.
%Finally, one may directly define the ground state as a state $\omega $ over $\cal A$ satisfying
%$\omega (A^*[H,A])\geq 0$ for all $A\in \cal A$ such that the commutator $[H,A]$ is defined
%(the so-called algebraic approach, see {\rm{\cite{BR2}}}).
%\end{remark}
%%%%%%%%%%%%%%%%%%%%%%%%%%%%%%%%%%%%%%%%%%%%%%%%%%%%%%%%%%%%%%%%%%%%%%%%%%%%%%%%%%%%%%%%%%%%%%%%%%%%%%%%%%%

Beside this introduction the paper consists of three sections. In
Section \ref{sec:2}, the procedure of elimination of the higher
components of the vector $F$ from Eq. (\ref{eigeneq}) and its
reduction to Eq. (\ref{effeq}) is presented in detail. Thereby we
consider directly the general case, where the components $f_n,\,
n>n_0$ with an arbitrary $n_0\geq 0$ are eliminated. In Sections
\ref{sec:3}, \ref{sec:4}, Eq. (\ref{effeq}) is analyzed for $n_0=0$ and $n_0=1$
and the invariant subspaces of the operator $H$ indicated above,
along with its (discrete and continuous) spectra in them, are constructed.

%%%%%%%%%%%%%%%%%%%%%%%%%%%%%%%%%%%%%%%%%%%%% Section 2 %%%%%%%%%%%%%%%%%%%%%%%%%%%%%%%%%%
\section{Reduction to a finite number of bosons}\label{sec:2}
\setcounter{equation}{0}

We shall show here how the spectral problem for $H$ can be reduced,
at sufficiently small coupling, to an equivalent problem within the
subspace with at most $n$ bosons.

It will be convenient to represent ${\cal H}$ as a space of
${\mathbb C}^2$-valued functions:
\begin{equation}\label{e1.1}
{\cal H}=L^2({\cal C},{\mathbb C}^2;d\mu ),
\end{equation}
where ${\cal C}=\bigcup _{n=0}^{\infty }{\cal C}_n$ is the set of
all finite subsets of ${\mathbb R}^d$, thereby $Q\in {\cal C}_n$ if
$|Q|=n$, and $d\mu $ is the so-called Lebesgue-Poisson measure:
\begin{equation}\label{e1.2}
d\mu (Q)=(1/|Q|!)\prod_{q\in Q}dq.
\end{equation}

Also, for $Q\in {\cal C}$, let $\omega (Q)=\sum _{q\in Q}\omega
(q)$. To simplify notation, we shall write $Q\setminus q:=Q\setminus
\{q\}$ and $Q\cup q:=Q\cup \{q\}$.

The Hamiltonian of the spin-boson model (\ref{Ham-sec-quant}) writes
in this representation
\begin{equation}
\label{e1.3}
\begin{array}{ll}
  (H_0F)(Q) & =[\varepsilon \sigma _3+ \omega (Q)]F(Q)\\ \\
(H_{int}F)(Q) & =\alpha \sigma _1[\sum _{q\in Q}\lambda
(q)F(Q\setminus q)+\int \overline{ \lambda (k)}F(Q\cup k)dk].
\end{array}
\end{equation}

Let us consider the orthogonal sum decomposition:
\begin{equation}\label{e1.4}
  {\cal H}={\cal H}_{\leq n}\oplus {\cal H}_{>n},
\end{equation}
where ${\cal H}_{\leq n}=\{ F\in {\cal H}; F(Q)=0, \forall Q, |Q|>n
\}\sim L^2(\bigcup _{k\leq n}{\cal C}_k,{\mathbb C}^2;d\mu )$.
Accordingly, the Hamiltonian (\ref{e1.3}) has a matrix
representation
\begin{equation}\label{e1.5}
  H=\begin{pmatrix}
    A_n & C_n \\
    C_n^* & B_n
  \end{pmatrix},
\end{equation}
where $A_n=P_{{\cal H}_{\leq n}}HP_{{\cal H}_{\leq n}}$, $B_n=P_{{\cal H}_{>
n}}HP_{{\cal H}_{>n}}$ (here, $P_{{\cal H}_{>n}},\, P_{{\cal H}_{>n}}$ denote the
orthogonal projections on the corresponding subspaces), and $C_n:{\cal
H}_{>n}\rightarrow {\cal H}_{\leq n}$ is given by
\begin{equation}\label{e1.6}
  (C_nF)(Q)=\delta _{|Q|,n}\alpha \sigma _1\int \overline{\lambda(k)}F(Q\cup
  k)dk,
\end{equation}
while $C_n^*:{\cal H}_{\leq n}\rightarrow {\cal H}_{>n}$ equals
\begin{equation}\label{e1.6a}
    (C_n^*F)(Q)=\delta _{|Q|,n+1}\alpha \sigma _1\sum\limits _{k\in
    Q}F(Q\setminus k)\lambda (k).
\end{equation}
As ${\rm Ran}(C_n)\subset {\cal H}_{n}$, $C_n^*$ can be viewed as an
operator $:{\cal H}_{n}\rightarrow {\cal H}_{>n}$.

The restriction to ${\cal H}_{\leq n}$ of the resolvent of the
Hamiltonian is obtained by solving for $F_n\in {\cal H}_{\leq n},
{\tilde F}_n\in {\cal H}_{>n}$ the system of two equations, where
$G\in {\cal H}_{\leq n}$:
\begin{equation}\label{e1.7}
  \left\{
  \begin{array}{rcrl}
    (A_n-z\mathbb{I}_{\leq n})F_n & + & C_n{\tilde F}_n & =G \\
     C_n^*F_n & + &
     (B_n-z\mathbb{I}_{>n}){\tilde F}_n & =0,
  \end{array}
  \right.
\end{equation}
where $\mathbb{I}_{\leq n},\mathbb{I}_{>n}$ are the unit operators in ${\cal H}_{\leq
n},{\cal H}_{>n}$, respectively.

For $z\in {\mathbb C}\setminus {\rm spec}(B_n)$, the second
Eq.(\ref{e1.7}) can be solved for ${\tilde F}_n$. Upon insertion of
the solution into the first Eq.(\ref{e1.7}), one obtains a reduced
problem in ${\cal H}_{\leq n}$:
\begin{equation}\label{e1.8}
(A_n-C_n(B_n-z\mathbb{I}_{>n})^{-1}C_n^*-z)F_n=G.
\end{equation}

If the operator in the l.h.s. is invertible and $F_n(z)$ is the solution of
Eq.(\ref{e1.8}), then $(F_n(z),\tilde F_n(z))$, where
\begin{equation}\label{e1.9}
\tilde F_n(z)=-(B_n-z\mathbb{I}_{>n})^{-1}C_n^*F_n(z),
\end{equation}
is the unique solution of Eq.(\ref{e1.7})

Our next task is to obtain a good characterization of the operator
\begin{equation}\label{Mn}
    M_n(z):=(B_n-z\mathbb{I}_{>n})^{-1}C_n^*:{\cal H}_n\rightarrow {\cal H}_{>n}.
\end{equation}

Let $\eta >0$, and define
\begin{equation}\label{Dn,eta}
    D_{n,\eta }:= \left\{z\in
\mathbb{C}:\,\, \textrm{Re}{\,z}<\lambda _{0,n+1}^0-\eta \right\}.
\end{equation}
(We remind that $\lambda _{0,n}^0:=-\varepsilon +n\kappa $ is the
first threshold of the $n$-boson branch of $H_0$.)

\begin{lemma}\label{L2.1}
There exists $\alpha _0=\alpha _0(n,\eta )$, such that, for any $\alpha <\alpha _0$
and $z\in D_{n,\eta }$, the operator $M_n(z)$, Eq. (\ref{Mn}), is bounded.
\end{lemma}
{\it Proof.} We represent $B_n$ in the form $$B_n=B_n^0+V_n,$$ where

$$(B_n^0F)(Q)=(\varepsilon \sigma _3+\omega (Q))F(Q),$$

$$(V_nF)(Q)=\alpha \sigma _1[\sum\limits _{k\in Q}\lambda (k)F(Q\setminus k)
+\int \overline{\lambda (k)}F(Q\cup k)dk], \; F\in {\cal H}_{>n}.$$
Hence, $M_n(z)$ is (formally) represented as
\begin{equation}\label{Mna}
    M_n(z)=(\mathbb{I}_{>n}+(B_n^0-z\mathbb{I}_{>n})^{-1}
V_n)^{-1}(B_n^0-z\mathbb{I}_{>n})^{-1}C_n^*.
\end{equation}

The assertion of Lemma \ref{L2.1} follows from the representation
(\ref{Mna}), if we prove that:
\begin{enumerate}
                       \item\label{L1-1} $\| (B_n^0-z\mathbb{I}_{>n})^{-1}
V_n\| _{{\cal H}_{>n}}<1$ for $z\in D_{n,\eta }$ and $\alpha $
sufficiently small;
                       \item\label{L1-2} $\| B_n^0-z\mathbb{I}_{>n})^{-1}
C_n^*\| _{{\cal H}_{>n}}<\infty $.
\end{enumerate}

To prove \ref{L1-1} we split $(B_n^0-z\mathbb{I}_{>n})^{-1} V_n$
into the sum of two terms:
$$\begin{array}{ccl} (S_1(z)F)(Q) & = & \left \{ \begin{array}{ll}
                  (\varepsilon \sigma _3+\omega(Q)-z)^{-1}\alpha \sigma _1
                  \sum\limits _{q\in Q}F(Q\setminus q)\lambda (q),
                  & {\rm if }\; |Q|>n+1 \\
                  0, & {\rm if}\; |Q|=n+1
                \end{array}\right. \\ \\ & & \\
(S_2(z)F)(Q) & = & (\varepsilon \sigma _3+\omega(Q)-z)^{-1}\alpha
\sigma _1\int \overline{\lambda (k)}F(Q\cup k)dk
\end{array}
$$
and estimate separately their norms.

Let $F_l\in L_2(\mathbb{R}^{dl})$ be the components of $F$. We have, for $l\geq n+2$,
$$ \| (S_1(z)F)_l\| _{L_2(\mathbb{R}^{dl})}<\frac{l\alpha \| \lambda\| _
{L_2(\mathbb{R}^{d})}}{(l-n-1)\kappa +\eta}\| F_{l-1}\| _{L_2(\mathbb{R}^{d(l-1)})}.
$$ Hence, $$
\begin{array}{ccl}
  \| S_1(z)F\| _{{\cal H}_{>n}}^2 & = & \sum\limits _{l\geq n+1}\frac{1}{l!}
  \| (S_1(z)F)_l\| _{L_2(\mathbb{R}^{dl})}^2 \\ & & \\
   & \leq & \alpha ^2\| \lambda \| ^2 _
{L_2(\mathbb{R}^{d})}\sum\limits _{l\geq n+2}\frac{l}{((l-n-1)\kappa +\eta)^2}\frac{\|
F_{l-1}\| _{L_2(\mathbb{R}^{d(l-1)})}^2}{(l-1)!}
\\ & & \\ & \leq & \alpha ^2\| \lambda\| ^2_
{L_2(\mathbb{R}^{d})}\max\limits _{l\geq n+2}\frac{l }{((l-n-1)\kappa +\eta )^2}\| F\|
_{{\cal H}_{>n}}^2=\alpha ^2\frac{(n+2)\| \lambda\| ^2 _
{L_2(\mathbb{R}^{d})}}{(\kappa +\eta )^2}\| F\| _{{\cal H}_{>n}}^2.
\end{array}
$$ A similar calculation gives the following estimate of the second term: $$ \|
S_2(z)F\| _{{\cal H}_{>n}}^2\leq \alpha ^2\frac{(n+1)\| \lambda\| ^2 _
{L_2(\mathbb{R}^{d})}}{\eta ^2}\| F\| _{{\cal H}_{>n}}^2. $$ As a consequence, the
inequality in \ref{L1-1} holds for $z\in D_{n,\eta }$, if $$\alpha <\frac{1}{2\|
\lambda\| _ {L_2(\mathbb{R}^{d})}}\min\{ \frac{\kappa +\eta}{\sqrt{n+2}},\frac{\eta
}{\sqrt{n+1}}\} $$
Point \ref{L1-2} can be proved similarly.   \hfill $\square$

We shall show that, for $z\in D_{n,\eta}$ and for $\alpha $ sufficiently small,
$M_n(z)$, Eq.(\ref{Mn}) has a particular representation, which we now define. Let
${\cal M}_2$ be the space of square $2\times 2$ complex matrices with some norm
$|\cdot |$ (e.g. $|n|=\frac{1}{2}\max _{i,j}|n_{i,j}|$), and $h:{\mathbb
R}^d\rightarrow (0,1]$ be the continuous square-integrable function appearing in
assumption (A2).

\begin{definition}\label{d1}
An operator $M_n:{\cal H}_n\rightarrow {\cal H}_{>n}$ is said to
have a $h$-regular representation in terms of coefficient functions,
if there exist continuously differentiable ${\cal M}_2$-valued
functions
\begin{equation*}
    \mu _n^{(m)}(\cdot ;\cdot ;\cdot ):{\cal C}\times {\cal
C}_{n-m}\times ({\mathbb R}^d)^m\rightarrow {\cal M}_2,\,\,\,
m=0,1,...,n,
\end{equation*}
where $\mu _n^{(m)}(Q_1;\cdot ;\cdot)=0$ for $|Q_1|\leq m,$ and $\mu
_n^{(m)}(Q_1;Q_2;\cdot)=0$ for $Q_1\bigcap Q_2\neq \emptyset ,$
satisfying, for some $M>0$, the estimation
\begin{equation}\label{e1.9'}
\sup\limits _{Q_2\in {\cal C}_{n-m}}\max\limits _{|\alpha |\leq
1}|\partial ^{\alpha }\mu
_n^{(m)}(Q_1;Q_2;k_1,...,k_m)|<M\prod\limits _{q\in
Q_1}h(q)\prod\limits _{i=1}^mh(k_i),
\end{equation}
such that, for $f\in {\cal H}_n$ and $|Q|>n$,
\begin{equation}\label{e1.10}
\begin{array}{rr}
  [M_nf](Q)= & \sum\limits _{m=0}^n \sum\limits _{\hat Q\subset Q, |\hat Q|=n-m} \int \mu
_n^{(m)}(Q\setminus {\hat Q};\hat Q;k_1,...,k_m) \\ \\
   & \times f(\hat Q\cup \left\{ k_1,...,k_m\right\} )dk_1...dk_m.
\end{array}
\end{equation}
\end{definition}
Here and below, we use the notation: for a multiindex $\alpha =\{
\alpha ^i_k;\; k=1,...,n,\; i=1,...,d\}$ and $Q=\{ q_1,...,q_n\} \in
{\cal C}_n$,
\begin{equation*}
\begin{array}{rcl}
(\partial ^{\alpha }f)(Q) & = & \prod\limits_{i,k}\frac{\partial
^{\alpha_k^i} }{\partial (q_k^i)^{\alpha _k^i}}f(Q), \\ & &
\\ |\alpha | & = & \sum\limits _{i,k}\alpha _k^i.
\end{array}
\end{equation*}

In Eq. (\ref{e1.10}), the sum over $\hat Q $ is a symmetrization, so
that the l.h.s. depends only on the set $Q$. As $f$ is permutation
symmetric, it is not necessary to impose the symmetry of $\mu
_n^{(m)}$ with respect to the $k$'s. The estimation (\ref{e1.9'})
means that the function $\mu _n^{(m)}$ and its gradient are bounded
uniformly, in particular $M$ is independent of $|Q_1|$. The set of
all coefficient functions for a given $h$ and $n$ is a Banach space
${\cal B}_n$ with the norm $\| \mu _n\|=\inf M$, where the infimum
is over all $M$ for which Eq.(\ref{e1.9'}) holds.

In order to show that $M_n(z)$ has a regular representation for all
$z\in D_{n,\eta }$ and determine the corresponding functions $\mu
_n^{(m)}(z)$, we shall use the identity
\begin{equation}\label{e1.12}
[1+(B_n^{(0)}-z\mathbb{I}_{>n})^{-1}V_n]M_n(z)=(B_n^{(0)}-z\mathbb{I}_{>n})^{-1}C_n^*.
\end{equation}

Suppose an operator $M_n:{\cal H}_n\rightarrow {\cal H}_{>n}$ is defined as in
Eq.(\ref{e1.10}) by the coefficient functions $\mu _n=\{ \mu
_n^{(m)}(Q_1;Q_2;k_1,...,k_m)\} $, where $|Q_2|=n-m$. Then, one can easily see that
$(B_n^{(0)}-z\mathbb{I}_{>n})^{-1}V_nM_n$ is likewise defined by a sequence of
coefficient functions, $\{ [\Gamma (z)\mu _n]^{(m)},\; m=0,...,n\} $, with
\begin{equation}\label{e1.13}
\begin{array}{ll}
[\Gamma (z)\mu_n]^{(m)}(Q_1;Q_2;k_1,...,k_m) & =\alpha (\varepsilon
\sigma _3 +\omega (Q_1\cup Q_2)-z)^{-1}\sigma _1 \\
 & \times \left\{ \sum\limits _{q\in Q_1}\lambda (q)\mu
 _n^{(m)}(Q_1\setminus q;Q_2;k_1,...,k_m)\right. \\ & +\int {\bar
 \lambda}(q')\mu
 _n^{(m)}(Q_1\cup q';Q_2;k_1,...,k_m)dq' \\  & \\
 & \left. + {\bar \lambda}(k_m)\mu
 _n^{(m-1)}(Q_1;Q_2\cup k_m;k_1,...,k_{m-1})\right\},
\end{array}
\end{equation}
where the last term does not appear if $m=0$. Let us also note that
the r.h.s. of Eq.(\ref{e1.12}) allows the representation
Eq.(\ref{e1.10}) with the coefficient functions $\hat \mu _n(z)$:
\begin{equation}\label{e1.14}
    [\hat \mu _n(z)]^{(m)}(Q_1,Q_2;k_1,..,k_m)=\delta _{m,0}\delta _{|Q_1|,1}\alpha
    [\varepsilon \sigma _3 +\omega (Q_2\cup q_1)-z]^{-1}\sigma _1\lambda
    (q_1),
\end{equation}
where we put $Q_1=\{ q_1\}$.  Therefore, Eq.(\ref{e1.12}) can be written as a
fixed-point equation for the coefficients $\mu _n=\mu _n(z)$:
\begin{equation}\label{e1.15}
\mu _n+ \Gamma (z)\mu _n= \hat \mu _n(z).
\end{equation}

%%%%%%%%%%%%%%%%%%%% Prop.t1 %%%%%%%%%%%%%%%%%%%%%%%%%%%%%%%%%%%%%%%%%%%%%%%%%%%%%%%%%%%%%%

\begin{proposition}\label{t1}
For any $\eta >0$ and $n\geq 0$, there exists $\alpha _0(\eta
,n)>0$, such that for any $\alpha <\alpha _0(\eta ,n)$, and for all
$z\in D_{n,\eta }$, Eq. (\ref{e1.15}) has a unique solution $\mu
_n(z)\in {\cal B}_n$, which is ${\cal B}_n$-valued analytic in
$D_{n,\eta }$ and $\| \mu _n(z)\| \leq K\alpha /\eta $ for some
constant $K$. Moreover,
\begin{equation}\label{mun_mum}
\mu_n^{(m)}(z;Q_1;Q_2;k_1,...,k_m)=\mu_m^{(m)}(z-\omega(Q_2);Q_1;\emptyset
;k_1,...,k_m)
\end{equation}
\end{proposition}
{\it Proof.} As both sides of Eq. (\ref{e1.13}) defining $\Gamma
(z)$ contain the same $Q_2$ and the same $m$-tuple $(k_1,...,k_m)$,
it will be convenient to treat these variables as parameters, define
(for $|Q_2|=n-m$) the functions of $Q_1$:
\begin{equation}\label{nu}
\nu
^{(m)}_{Q_2,(k_1,...,k_m)}(Q_1):=\mu_n^{(m)}(Q_1;Q_2;k_1,...,k_m),
\end{equation}
and remark that
\begin{equation}\label{e1.13'}
\begin{array}{l}
[\Gamma (z)\mu_n]^{(m)}(Q_1;Q_2;k_1,...,k_m)  =  [\Delta _m(z-\omega (Q_2))\nu
^{(m)}_{Q_2,(k_1,...,k_m)}](Q_1) \\  \\ +\alpha [\varepsilon \sigma _3 +\omega
(Q_1)-(z-\omega (Q_2))]^{-1} \cdot \sigma _1\lambda (k_m)\nu ^{(m-1)}_{Q_2\cup
k_m,(k_1,...,k_{m-1})}(Q_1),
\end{array}
\end{equation}
where we used that $\omega (Q_1\cup Q_2)=\omega (Q_1)+\omega (Q_2)$
for $Q_1\cap Q_2=\emptyset $. The operator $\Delta _m(z),\, z\in
D_{m,\eta }$ acts on the functions $\nu (Q)$ (such that $\nu (Q)=0$
 for $|Q|\leq m)$ according to the formula
\begin{equation}\label{Delta}
\begin{array}{ccl}
  (\Delta _m(z)\nu )(Q) & = & \chi _{>m}(Q)\alpha (\varepsilon \sigma _3 +\omega (Q)-z)^{-1} \\
  & & \\
   & & \times \sigma _1\{ \sum\limits _{q\in Q}\lambda (q)\nu (Q\setminus q)+
   \int \overline{\lambda (k)}\nu (Q\cup k)dk\} ,
\end{array}
\end{equation}
where $\chi _{>m}(Q)$ is the indicator of the set ${\cal C}_{>m}$.

Let us consider the Banach space $\tilde {\cal B}$ of all
continuously differentiable functions $\{ \nu :{\cal C}\rightarrow
{\cal M}_2\}$, for which
\begin{equation}\label{e1.19}
  \| \nu \|=\sup\limits _{Q\in {\cal C}}\frac{\max\limits _{|\alpha |\leq 1}|\partial ^{\alpha }
  \nu (Q)|}{\prod\limits _{q\in Q}h(q)}<\infty ,
\end{equation}
where $|\cdot |$ denotes the norm in ${\cal M}_2$. Also, let $\tilde {\cal
B}_{>m}\subset \tilde {\cal B}$ be the subspace of functions $\nu $ which vanish on
${\cal C}_{\leq m}$, i.e. $\nu (Q)=0,\,\forall Q, |Q|\leq m$.

%%%%%%%%%%%%% Lemma 2.4 %%%%%%%%%%%

\begin{lemma}\label{L1.2}
For every $\eta >0$ and $m\geq 0$, there exists ${\hat \alpha
}_0(\eta ,m)>0$, such that for any $\alpha <{\hat \alpha }_0(\eta
,m)$, $\Delta _m(z)$ is a bounded operator in $\tilde {\cal
B}_{>m}$, norm-analytic of $z$ in $D_{m,\eta }$, and
\begin{equation}\label{e1.21}
  \sup\limits _{z\in D_{m,\eta }}\| \Delta _m(z)\| \leq 1/2.
\end{equation}
Also, the function $$\hat \nu (z-\omega (Q_2);\cdot ):={\hat \mu }^{(0)}_n(z;\cdot
;Q_2;\emptyset )= {\hat \mu }^{(0)}_0(z-\omega (Q_2);\cdot ;\emptyset ;\emptyset )$$
is a $\tilde {\cal B}_{>0}$-valued analytic function of $z\in D_{n,\eta }$ $(n=|Q_2|)$
and $\| \hat \nu (z-\omega (Q_2))\| \leq C\alpha /\eta $ in $D_{n,\eta }$ for a
certain constant $C$.
\end{lemma}

The proof of this lemma is similar to that of Lemma
\ref{L2.1}.\hfill $\square$

%%%%%%%%%%%%%%%%%% Remark-new %%%%%%%%%%%%%%%%%%%%%%

\begin{remark}\label{new}
We shall consider also the Banach space $\tilde {\tilde {\cal B}}$ consisting of
continuously differentiable functions $\nu :{\cal C}\rightarrow \mathbb{C}^2$ with the
norm $\| \nu \|$ given by Eq. (\ref{e1.19}), in which now $|\cdot |$ means the usual
norm on $\mathbb{C}^2$. Its subspaces ${\tilde {\tilde {\cal B}}}_n, {\tilde {\tilde
{\cal B}}}_{\leq n}, {\tilde {\tilde {\cal B}}}_{>n}$ are introduced as above. It
follows from Proposition \ref{t1} and the representation (\ref{e1.10}) that, for all
$z\in D_{n,\eta }$, $M_n(z)$ applies the space ${\tilde {\tilde {{\cal B}'}}}_n$, the
dual space of ${\tilde {\tilde {\cal B}}}_n$, into ${\tilde {\tilde {{\cal
B}'}}}_{>n}\subset {\cal H}_{>n}$ and is bounded with respect to the norm of ${\tilde
{\tilde {{\cal B}'}}}$. Besides, assumptions (A1) and (A2) imply that the
operator $H$ can be extended to an unbounded operator acting in ${\tilde {\tilde
{{\cal B}'}}}$ (denoted also $H$) on a domain including all finite (with respect to
the spatial variables, as well as to the number of variables) elements of
${\tilde{\tilde {{\cal B}'}}}$:
\begin{equation}\label{dual space B}
\{ \phi\in {\tilde {\tilde {{\cal B}'}}}: \phi (Q)=0,\; {\rm
if}\; \exists R, \exists N,\; {\rm dist}(0,Q)>R\; {\rm or}\; |Q|>N\}
\end{equation}
\end{remark}

%%%%%%%%%%%%%%%%%%%%%%%%%%%%%%%%%%%%%%%%%%%%%%%%%%%%%%%%

\medskip\
In this way, as follows from Eq. (\ref{e1.13'}), $\nu
^{(0)}_{Q_2,\emptyset }(\cdot )$ satisfies the equation
(\ref{e1.15}) for $m=0$, which writes as
%%\begin{equation}\label{m=0}
%    \nu ^{(0)}_{Q_2,\emptyset }(\cdot )+
%    \Delta _0(z-\omega (Q_2))\nu ^{(0)}_{Q_2,\emptyset }(\cdot )
%    ={\hat \nu }(z-\omega (Q_2))(\cdot ).
%\end{equation}
\begin{equation}\label{m=0}
    \nu (\cdot )+
    \Delta _0(z-\omega (Q_2))\nu (\cdot )
    ={\hat \nu }(z-\omega (Q_2);\cdot ).
\end{equation}
If $z\in D_{n,\eta }$, the difference $z-\omega (Q_2)$ belongs to
$D_{0,\eta }$, therefore, according to Lemma \ref{L1.2}, for $\alpha
<\alpha _0(\eta ,0)$, the equation has one solution
%\begin{equation}\label{solm=0}
%    \nu ^{(0)}_{Q_2,\emptyset }(z-\omega (Q_2))(\cdot )\in {\tilde
%    {\cal B}}_{>0},
%\end{equation}
\begin{equation}\label{nu00}
    \nu ^{(0)}_{Q_2,\emptyset }(z;\cdot )=\nu ^{(0)}_{\emptyset ,\emptyset }
    (z-\omega (Q_2);\cdot )\in {\tilde {\cal B}}_{>0},
\end{equation}
where the equality comes from the fact that both sides obey the same equation
(\ref{m=0}), therefore both equal its unique solution,  $ (\mathbb{I}+\Delta
_0(z-\omega (Q_2))^{-1}{\hat \nu }(z-\omega (Q_2))$. Obviously, the solution is
analytic of $z\in D_{n,\eta }$ and (by the smoothness of $\omega $) continuously
differentiable of $Q_2=\{ q_1,...,q_n\} $. For its norm  we have, by Lemma \ref{L1.2},
the estimate:
%\begin{equation}\label{normm=0}
%    \| \nu ^{(0)}_{Q_2,\emptyset }\| <2C\alpha /\eta .
%\end{equation}
\begin{equation}\label{normm=0}
    \sup\limits _{z\in D_{n,\eta },Q_2\in {\cal C}_n}\| \nu ^{(0)}_{Q_2,\emptyset }(z)\|
   \leq \sup\limits _{z\in D_{0,\eta}}
   \| (\mathbb{I}+\Delta _0(z))^{-1}\| \| {\hat \nu }(z)\|
    <2C\alpha /\eta .
\end{equation}

We consider next the case $m>0,\, m\leq n$. Eq. (\ref{e1.15})
satisfied by $\nu ^{(m)}_{Q_2,(k_1,..,k_m)}(\cdot )$ writes
%\begin{equation}\label{m>0}
%\begin{array}{l}
%  \nu ^{(m)}_{Q_2,(k_1,..,k_m)}(\cdot )+
%  \Delta_m(z-\omega (Q_2))\nu ^{(m)}_{Q_2,(k_1,..,k_m)}(\cdot ) \\
%  =\alpha [\varepsilon \sigma _3 +\omega (\cdot )-(z-\omega (Q_2))]^{-1}
% \sigma _1\lambda (k_m)\nu ^{(m-1)}_{Q_2\cup
%k_m,(k_1,...,k_{m-1})}(\cdot ),
%\end{array}
%\end{equation}
\begin{equation}\label{m>0}
\begin{array}{l}
  \nu (\cdot )+
  \Delta_m(z-\omega (Q_2))\nu (\cdot ) \\
  =-\alpha [\varepsilon \sigma _3 +\omega (\cdot )-(z-\omega (Q_2))]^{-1}
 \sigma _1\lambda (k_m)\nu ^{(m-1)}_{Q_2\cup
k_m,(k_1,...,k_{m-1})}(z;\, \cdot \, ),
\end{array}
\end{equation}
and, for $\alpha <\alpha _0(\eta ,m)$, allows to determine
inductively $\nu ^{(m)}_{Q_2,(k_1,...,k_m)}(z;\cdot )$ in terms of
the solution $\nu ^{(m-1)}_{Q_2\cup k_m,(k_1,...,k_{m-1})}(z; \cdot
)$ of the ($m-1$)th equation. Suppose that $\nu
^{(m-1)}_{Q'_2,(k_1,...,k_{m-1})}(z)(\cdot )$ has been shown to
fulfill:
\begin{enumerate}
    \item For $z\in D_{|Q'_2|+m-1,\eta}$,
    \begin{equation}\label{yyy}
        \nu ^{(m-1)}_{Q'_2,(k_1,...,k_{m-1})}(z)=
    \nu ^{(m-1)}_{\emptyset ,(k_1,...,k_{m-1})}
    (z-\omega (Q'_2))\in {\cal B}_{>m-1},
    \end{equation}
    and
    \begin{equation}\label{normm-1}
        \| \nu ^{(m-1)}_{\emptyset ,(k_1,...,k_{m-1})}(z)\| _{{\tilde
    {\cal B}}_{>m-1}}<C2^m(\alpha /\eta )^m\prod\limits
    _{i=1}^{m-1}h(k_i);
    \end{equation}
%  \item $\nu ^{(m-1)}_{Q'_2,(k_1,...,k_{m-1})}(z)\in {\tilde
%    {\cal B}}_{>m-1}$ for $z\in D_{|Q'_2|+m-1,\eta}$ and
%    \begin{equation}\label{normm-1}
%        \| \nu ^{(m-1)}_{Q'_2,(k_1,...,k_{m-1})}(z)\| _{{\tilde
%    {\cal B}}_{>m-1}}<C2^m(\alpha /\eta )^m\prod\limits
%    _{i=1}^{m-1}h(k_i);
%    \end{equation}

  \item $\nu ^{(m-1)}_{\emptyset ,(k_1,...,k_{m-1})}(z)$ is a ${\tilde
    {\cal B}}_{>m-1}$-analytic function of $z\in
    D_{m-1,\eta}$, differentiable of $k_1,...,k_{m-1}$ and
    \begin{equation}\label{derivm-1}
        \sup\limits _{z\in D_{m-1,\eta}}\max\limits _{1\leq i\leq
        m-1}|\partial _{k_i}\nu ^{(m-1)}_{\emptyset ,(k_1,...,k_{m-1})}(z;Q)|
        \leq C'2^m(\alpha /\eta )^m\prod\limits
    _{i=1}^{m-1}h(k_i)\prod\limits
    _{q\in Q}h(q),
    \end{equation}
where the constants $C,C'$ do not depend on $m,k_1,...,k_{m-1},Q$.
%\item $\nu ^{(m-1)}_{Q'_2,(k_1,...,k_{m-1})}(z)$ is a ${\tilde
%    {\cal B}}_{>m-1}$-analytic function of $z\in
%    D_{|Q'_2|+m-1,\eta}$, differentiable of $k_1,...,k_{m-1}$ and
%    \begin{equation}\label{derivm-1}
%        \sup\limits _{z\in D_{|Q'_2|+m-1,\eta}}\max\limits _{1\leq i\leq
%        m-1}|\partial _{k_i}\nu ^{(m-1)}_{Q'_2,(k_1,...,k_{m-1})}(z)(Q)|
%        \leq C'2^m(\alpha /\eta )^m\prod\limits
%    _{i=1}^{m-1}h(k_i)\prod\limits
%    _{q\in Q}h(q),
%    \end{equation}
%where the constants $C,C'$ do not depend on
%$m,k_1,...,k_{m-1},Q'_2,Q$.
\end{enumerate}
Remark that, by Eqs. (\ref{nu00}) and (\ref{normm=0}),  $\nu
^{(0)}_{Q_2,\emptyset }(z)$ fulfills both conditions.

Using Lemma \ref{L1.2} we find that Eq. (\ref{m>0}) has, for $\alpha
<\alpha _0(\eta ,m)$ and $z\in D_{|Q_2|+m}$, one solution
\begin{equation}\label{xxx}
\begin{array}{l}
  \nu ^{(m)}_{Q_2,(k_1,...,k_{m})}(z;\cdot ) \\
  \hfill =(\mathbb{I}+\Delta_m(z-\omega (Q_2)))^{-1}
    \alpha [\varepsilon \sigma _3 +\omega (\cdot )-(z-\omega (Q_2))]^{-1}
 \sigma _1\lambda (k_m)\nu ^{(m-1)}_{Q_2\cup k_m,(k_1,...,k_{m-1})}(z;\cdot )
\end{array}
\end{equation}
which has the analogous properties. This proves the existence of the
coefficient functions $\mu_n^{(m)}(z;Q_1;Q_2;k_1,...,k_m)$ and their
estimates (\ref{e1.9'}).

The equality (\ref{yyy}) in the case $m>0$ follows again by
induction with respect to $m$:

Taking $Q_2=\emptyset $ in Eq. (\ref{xxx}), we have:
\begin{equation}\label{xx}
    \begin{array}{l}
  \nu ^{(m)}_{\emptyset ,(k_1,...,k_{m})}(z) \\
  \hfill =(\mathbb{I}+\Delta_m(z))^{-1}
    \alpha [\varepsilon \sigma _3 +\omega (\cdot )-z)]^{-1}
 \sigma _1\lambda (k_m)\nu ^{(m-1)}_{\{ k_m\},(k_1,...,k_{m-1})}(z;\cdot )
\end{array}
\end{equation}
Suppose that (\ref{yyy}) holds true; then,
$$\nu ^{(m-1)}_{Q_2\cup k_m,(k_1,...,k_{m-1})}(z;\cdot )
=\nu ^{(m-1)}_{\{ k_m\},(k_1,...,k_{m-1})}(z-\omega (Q_2);\cdot
)=\nu ^{(m-1)}_{\emptyset ,(k_1,...,k_{m-1})}(z-\omega (Q_2)-\omega
(k_m);\cdot ),$$ wherefrom it follows that the r.h.s. of Eq.
(\ref{xx}) written for $z\mapsto z-\omega (Q_2)$ coincides with the
r.h.s. of Eq. (\ref{xxx}). This proves (\ref{yyy}) for $m$, hence
the equality  (\ref{mun_mum}).

The proposition \ref{t1} is proved. \hfill $\square $

\medskip\
The following corollary collects the information on the structure of the operator
$C_n(B_n-z\mathbb{I}_{>n})^{-1}C_n^*$ acting in ${\cal H}_n$ implied by the regular
representation of $M_n(z)$.
\begin{corollary}\label{l1.3}
The following representation holds: for $\alpha <\alpha _0(\eta ,n)$, and $f\in {\cal
H}_n,\; Q\in {\cal C}_n$,
\begin{equation}\label{e1.26}
\begin{array}{l}
  [C_n(B_n-z\mathbb{I}_{>n})^{-1}C_n^*f](Q)=m_n(z;Q)f(Q)+ \\ \sum\limits _{m=1}^n
  \sum\limits _{Q_1\subset Q;|Q_1|=m}\int
  {\hat D}_{n,m}(z;Q_1;Q\setminus Q_1;k_1,...,k_m)f((Q\setminus Q_1)\cup\{
  k_1,...,k_m\})dk_1...dk_m,
  \end{array}
\end{equation}
where
\begin{enumerate}
  \item\label{cond-m} $m_n(z;Q)=m_0(z-\omega (Q);\emptyset )$ is analytic of
  $z\in D_{n,\eta }$, and
  $$|m_n(z;Q)|\leq 2C'(\alpha ^2/\eta)\|\lambda
  \|^2_{L_2(\mathbb{R}^d)};$$
  Hence, $m_n(z;Q)$ is continuously differentiable of $Q\in {\cal
  C}_n$, as well.
  \item\label{cond-D} ${\hat D}_{n,m}(z;Q_1;Q_2;k_1,...,k_m)=
  {\hat D}_{m,m}(z-\omega (Q_2);Q_1;\emptyset ;k_1,...,k_m)$
  is analytic of
  $z\in D_{n,\eta }$, continuously differentiable of $\{
  Q_1;Q_2;k_1,...,k_m\}\in
  {\cal C}_m\times {\cal C}_{n-m}\times (\mathbb{R}^d)^m$,  and
  \begin{equation}\label{e-cond-D}
   \max\limits _{|\alpha |\leq 1}|\partial ^{\alpha }{\hat D}_{n,m}(z;Q_1;Q_2;k_1,...,k_m)|
   \leq C'(2\alpha /\eta )^{m+1}
  \prod\limits _{q\in Q_1} h(q)\prod\limits _{i=0}^m h(k_i),
  \end{equation}
  where $C'$ is a constant.
\end{enumerate}
\end{corollary}
{\it Proof.} The assertions follow by applying the representation
(\ref{e1.10}), the identity Eq.(\ref{mun_mum}) and the estimates
(\ref{normm-1}), (\ref{derivm-1}) in the formulae:
\begin{equation}\label{e1.27}
m_n(z;Q)=\alpha \sigma _1 \int \overline{\lambda (q')}\mu _n^{(0)}(z;q';Q;\emptyset
)dq',
\end{equation}

\begin{equation}\label{e1.28}
\begin{array}{rl}
{\hat D}_{n,m}(z;Q_1;Q_2;k_1,...,k_m) & =\alpha \sigma _1\int \overline{\lambda
(q')}\mu _n^{(m)}(z;Q_1\cup q';Q_2;k_1,...,k_m)dq' \\ & \\ & +\alpha \sigma
_1\overline{\lambda (k_m)}\mu _n^{(m-1)}(z;Q_1;Q_2\cup k_m;k_1,...,k_{m-1}).
\end{array}
\end{equation}
\hfill $\square$

The ${\cal M}_2$-valued kernels ${\hat D}(Q_1;Q_2;k_1,...,k_m)$, which are
continuously differentiable on ${\cal C}_m\times (\mathbb{R}^d)^m$ and dominated by
$h$ as in Eq.(\ref{e-cond-D}), form a Banach space ${\cal K}_m$ with the norm $$\|
{\hat D}\|=\sup\limits _{Q_1,Q_2,k_1,...,k_m}\max\limits _{|\alpha |\leq 1}|\partial
^{\alpha }{\hat D}(Q_1;Q_2;k_1,...,k_m)|/\prod\limits _{q\in Q_1}h(q)\prod\limits
_{i=0}^m h(k_i),$$ where $|\cdot |$ is the norm in ${\cal M}_2$.

%%%%%%%%%%%%%%%%%%%%%%%%%%%%%%%%%%%%%%%%%%% Section 3 %%%%%%%%%%%%%%%%%%%%%%%%%%%%%%%%%%%%%%%%%%%%
\section{Discrete spectrum}\label{sec:3}
\setcounter {equation}{0}

We consider here, as an example of the general analysis, the cases
$n=0,1$. This allows the construction of the eigenvectors and of
part of the one-boson branch. In this section we consider the \textit{discrete part} of the spectrum.
\medskip\

{\bf I}. For $n=0$, the equation (\ref{e1.8}) becomes an equation in
$\mathbb{C}^2$:
\begin{equation}\label{e3.1}
  (\varepsilon \sigma _3-m_0(z;\emptyset )-z)f=g,
\end{equation}
where the matrix $m_0(z;\emptyset )$ is analytic in $D_{0,\eta }$
and $| m_0(z;\emptyset )|\leq 2C'\| \lambda \| ^2\alpha ^2/\eta $.
Eq.(\ref{e3.1}) has a unique solution unless $z$ is a (real) zero of
the determinant of the matrix in the l.h.s., i.e., denoting $m(z):=m_0(z;\emptyset )$,
\begin{equation}\label{e3.1a}
\det(\varepsilon \sigma _3-m(z)-z)=0.
\end{equation}
This equation can be brought to the form
\begin{equation}\label{e3.2}
  1+\frac{m(z)_{11}+(1/2\varepsilon ){\rm det}(m(z))}{z-\varepsilon }+
  \frac{m(z)_{22}-(1/2\varepsilon ){\rm det}(m(z))}{z+\varepsilon }=0.
\end{equation}

Now, $m(z)=C_0(B_0-z)^{-1}C_0^{*}$ is positive and increasing for $z\in (-\infty
,\lambda ^0_{0,1}-\eta )$. As $0<m(z)_{ii}=0(\alpha ^2)$, while $0<{\rm
det}(m(z))=0(\alpha ^4)$, both numerators are positive at $z=\mp \varepsilon $. The
graph of the function in the l.h.s. is schematically depicted in Fig.1 (for the case
$\varepsilon <\lambda _{0,1}^0-\eta $).

% \begin{center}
% %   % Requires \usepackage{graphicx}

\includegraphics{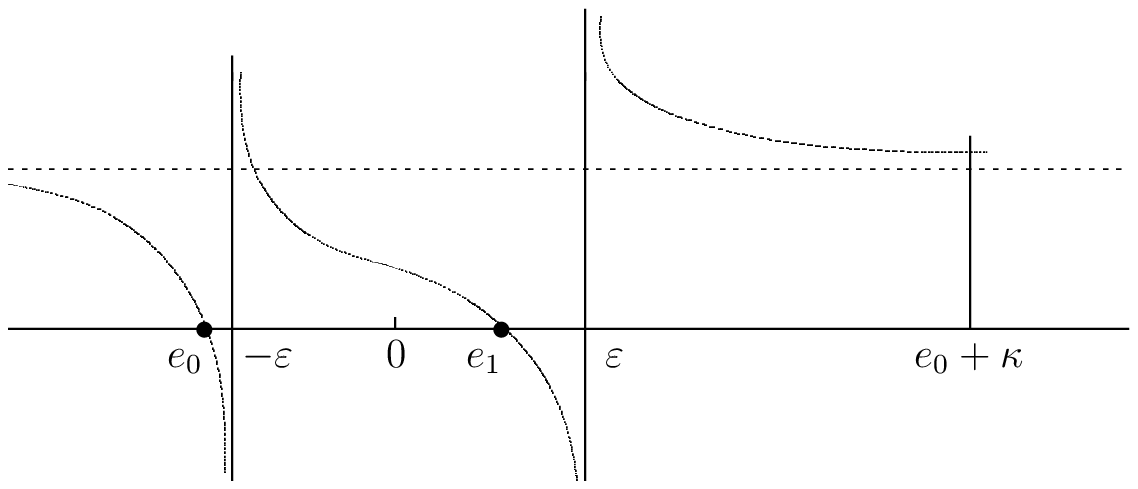}

\begin{center}
  \footnotesize{Figure 1a: The graph of the l.h.s.\ of (\ref{e3.2}) : case of two roots}
\label{fig1a}
\end{center}

\includegraphics{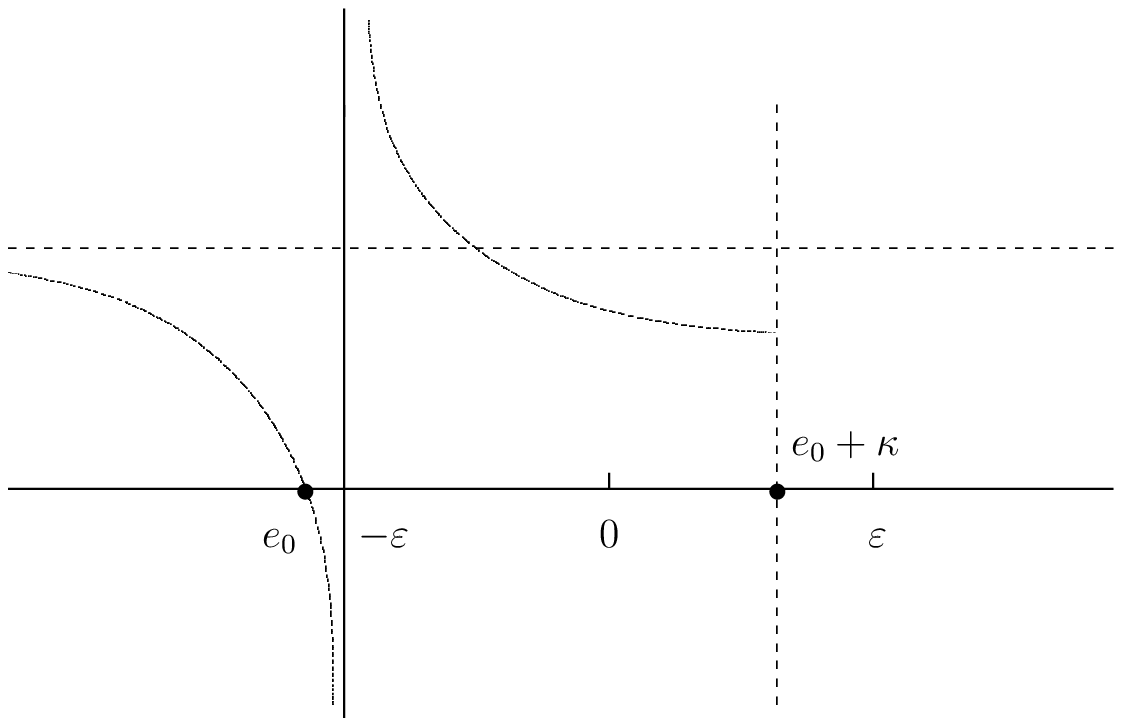}

\begin{center}
  \footnotesize{Figure 1b: The graph of the l.h.s.\ of (\ref{e3.2}) : case of one root}
\label{fig1b}
\end{center}

One concludes that there exists always a zero $e_0<-\varepsilon $. Also, whenever
$\varepsilon <\lambda_{0,1}^0-\eta =-\varepsilon +\kappa -\eta $, there exists a
second zero $-\varepsilon <e_1<\varepsilon $. For $z=e_i,\,\, i=0,1$, Eq.(\ref{e3.1})
with $g=0$ has nontrivial normalized solutions $f_i$, which can be completed with the
higher components $-M_0(e_i)f_i$ to eigenvectors of $H$:
\begin{equation}\label{e3.2a}
    F_0^{(i)}(Q)=\left\{
    \begin{array}{ll}
    f_i & ,\; Q=\emptyset \\ (-M_0(e_i)f_i)(Q) & ,\; Q \neq \emptyset .
    \end{array}
    \right.
\end{equation}
Eq.(\ref{e3.1a}) has no other real solution $z<\lambda_{0,1}^0-\eta $. Also, there are
no complex solutions ${\tilde z},\, \textrm{Im} {\tilde z}\neq 0$: otherwise, considering an
eigenvector $\tilde f$ of the matrix $\varepsilon \sigma _3-m(\tilde z)$ corresponding
to the eigenvalue $\tilde z$ and completing it with the vector $-M_0(\tilde z)\tilde
f\in {\cal H}_{>0}$, one would obtain an eigenvector of $H$ with a non-real
eigenvalue.
\medskip\

>From now on, we consider for definiteness the case in which Eq. (\ref{e3.2}) has two
solutions $e_0,e_1<\lambda ^0_{0,1}-\eta $. The case with one solution is treated
similarly.
\medskip\

{\bf II}. For $n=1$, the equation (\ref{e1.8}) is a system of two equations, valid for
$z\in D_{1,\eta }$:
\begin{equation}\label{e3.3}
    \left\{ \begin{array}{rcrcl}
             (\varepsilon \sigma _3-z)f_0 & + &
             \alpha \sigma _1\int {\bar \lambda }(k)f_1(k)dk\hfill
             & = & g_0 \\ & & & & \\
             \alpha \sigma _1\lambda (q)f_0 & + & [\varepsilon \sigma _3-(z-\omega (q))
             -m(z-\omega (q))]f_1(q) & &  \\ & & & & \\
              & & -\int {\hat D}_{1,1}(z;q;\emptyset ;k)f_1(k)dk & = & g_1(q)
           \end{array}
    \right.
\end{equation}

The operator
\begin{equation}\label{e3.4}
    [B(z)f_1](q)=[\varepsilon \sigma _3-z+\omega (q)
    -m(z-\omega (q))]f_1(q)-\int {\hat D}_{1,1}(z;q;\emptyset
    ;k)f_1(k)dk
\end{equation}
has an analytic inverse whenever the matrices $\varepsilon \sigma
_3-(z-\omega (q)) -m(z-\omega (q)),\,\, \forall q\in \mathbb{R}^d$
are invertible, what happens for $z\in \mathbb{C}\setminus I$, where
we denoted $I=[e_0+\kappa ,\infty )$. As we shall see below,
${\lambda }_{0,1}=e_0+\kappa $ is the left boundary of the spectrum
of $H$. Let ${\bar D}_0=\{ z\in \mathbb{C}: \textrm{Re}{\, z}<{\lambda }_{0,1}
\},\; n=1,2,...$. The following proposition holds:
%%%%%%%%%%%%%%%%%%%%%%%%%%%%%%%%%%%%%%%%%%%%%%% Propositio 3.1 %%%%%%%%%%%%%%%%%%%%%%%%%%%%%%%%%%%%%%%%%%%%%%
\begin{prop}\label{P3.1}
For $\eta >0$ sufficiently small, $\alpha <\alpha _0(\eta , 1)$ and
$z\in D_{1,\eta }\setminus I$ the inverse $B(z)^{-1}$ exists and has
the representation
\begin{eqnarray}\label{e3.5}
[B(z)^{-1}f](q) &=& [\varepsilon \sigma_3-(z-\omega (q)) - m(z-\omega (q))]^{-1} \\
&\times & \{ f(q)+\int K(z;q,k) [\varepsilon \sigma_3-(z-\omega (k)) -m(z-\omega (k))]^{-1}f(k) \, dk\} \ , \nonumber
\end{eqnarray}
where the kernel $K(z;\cdot ,\cdot )\in {\cal K}_1$ and its ${\cal
K}_1$-norm is uniformly bounded for $z\in {\bar D}_{0}$. Besides,
$K(z;\cdot ,\cdot )$ is a ${\cal K}_1$-valued analytic function of
$z$ in $D_{1,\eta }\setminus I$, which has boundary values at the
cut $I$, i.e. for all $x\in [e_0+\kappa ,\lambda ^0_{0,2}-\eta )$
the following limits exist in ${\cal K}_1$:
\begin{equation}\label{e3.6}
K^\pm (x;\cdot ,\cdot )= \lim\limits _{\epsilon \searrow 0}K(x\pm
i\epsilon;\cdot ,\cdot ).
\end{equation}
The kernels $K^\pm (x;\cdot ,\cdot )$ are H\"{o}lder-continuous
${\cal K}_1$-valued functions of $x\in [e_0+\kappa ,\lambda
^0_{0,2}-\eta )$.
\end{prop}
%%%%%%%%%%%%%%%%%%%%%%%%%%%%%%%%%%%%%%%%%%%%%%%%%%%%%%%%%%%%%%%%%%%%%%%%%%%%%%%%%%%%%%%%%%%%%%%%%%%%%%%%%%%%%%%%

{A similar statement is outlined in our paper \cite{AMZ}. For the reader convenience below we shortly resume
the proof of the Proposition \ref{P3.1} and we leave a detailed demonstration (which includes a generalization
of the Privalov lemma) for \textit{Appendix}. Note that essentially the proof consists in the following :}\\
%%%%%%%%%%%%%%%%%%%%%%%%%%%%%%%%%%%%%%%%%%%%%%%%%%%%%%%%%%%%%%%%%%%%%%%%%%%%%%%%%%%%%%%%%%%%%%%%%%%%%%%%%%%%%%%%%
Denoting by ${\hat D}_{1,1}$ the integral operator with the kernel ${\hat
D}_{1,1}(z;q;\emptyset ;k)\in {\cal K}_1$, one has:
\begin{equation}\label{e3.4'}
\begin{array}{lll}
B(z)^{-1} & = & (\varepsilon \sigma _3-(z-\omega (\cdot ))
-m(z-\omega (\cdot )))^{-1} \\ & + & \{ \sum\limits _{k=1}^{\infty
}[(\varepsilon \sigma _3-(z-\omega (\cdot )) -m(z-\omega (\cdot
)))^{-1}{\hat D}_{1,1}]^k\} \\ & \times & (\varepsilon \sigma
_3-(z-\omega (\cdot )) -m(z-\omega (\cdot )))^{-1},
\end{array}
\end{equation}
where every term of the sum is an integral operator with kernel in
${\cal K}_1$ and the sum converges in ${\cal K}_1$ for $\alpha $
sufficiently small, uniformly for $z\in D_{1,\eta }$. Hence we
arrive at the representation (\ref{e3.5}), where the kernel
$K(z;\cdot ,\cdot )\in {\cal K}_1$ depends analytically on $z\in
D_{1,\eta }\setminus I$. The existence of the limits (\ref{e3.6})
and the properties of the boundary value kernels $K^\pm (x;\cdot
,\cdot )$ are proved for every term of the series (\ref{e3.4'})
using induction over $k$. Thereby, we use that, if two $z$-dependent
kernels $K_1(z),K_2(z)\in {\cal K}_1$ possess boundary values like
in Eq.(\ref{e3.6}), then the kernel
\begin{equation*}
K_3(z;q,k): =\int K_2(z;q,q') [\varepsilon \sigma _3-(z-\omega (q')) -m(z-\omega (q'))]^{-1} K_1(z;q',k)\ dq' \ ,
\end{equation*}
has the same property, in view of the
\textit{Sokhotski formula}:  $1/(x+i0)={\cal P}(1/x)+i\pi \delta(x)$. Indeed,
the inverse matrix $(\varepsilon \sigma _3-(z-\omega (q'))
-m(z-\omega (q')))^{-1}$ has the structure: either
$$\frac{A_0}{e_0-(z-\omega (q'))}+\phi (z-\omega (q')),$$ if Eq.
(\ref{e3.1a}) has one solution $e_0$; or $$\frac{A_0}{e_0-(z-\omega
(q'))}+\frac{A_1}{e_1-(z-\omega (q'))}+\phi (z-\omega (q')),$$ if a
second solution $e_1$ exists, too. Here, $A_0,A_1$ are $2\times
2$-matrices and $\phi (z)$ is a ${\cal M}_2$-valued analytic
function in ${D}_{0,\eta }$. Therefore, denoting by $d\nu _x(q)$ the
\textit{Gelfand-Leray measure} on the surface ${\cal C}_{1,y}=\{ q':\; \omega
(q')=y\}$, we obtain
\begin{equation*}\begin{array}{ccl}
                   K_{\pm }(x;q,k) & = & \int _{\kappa
}^{\infty }dy \left[ (\sum\limits _{j=0,1} A_j/[{e_j+y-x\pm
i0}]+\phi (x-y)) \right. \\
                    & & \times \left. \int _{{\cal C}_{1,y}}K_{1,\pm }(x;q;q')K_{2,\pm
}(x;q';k)d\nu _y(q')\right] dx.
                 \end{array}
\end{equation*}
As the internal integral over the surface ${\cal C}_{1,y}$ is a smooth function of
$y$, the integral with respect to $y$ can be done and gives H\"{o}lder continuous
functions of $x$, $K_{i,\pm }(x;\cdot ,\cdot),\, i=1,2$ (as follows from the
Plemelj-Privalov theorem  \cite{Mus}, \cite{Pr}, and Appendix).

%%%%%%%%%%%%%%%%%%%%%%%%%%%%%%%%%%%%%%%%%%%%%%%%%%%%%%%%%%%%%%%%%%%%%%%%%%%%%%%%%%%%%%%%%%%%%
Next, we consider Eq.(\ref{e3.3}) for $g_1=0$, i.e. again the
resolvent of $H$ restricted to ${\cal H}_0$. We have $$f_1(\cdot
)=-\alpha (B(z))^{-1}\sigma _1\lambda (\cdot )f_0.$$ Plugging $f_1$
into the first equation (\ref{e3.3}) we obtain that
$$\left( \varepsilon \sigma _3-z-\alpha ^2\sigma _1
\int {\bar \lambda }(q)[B(z)^{-1}\sigma _1\lambda(\cdot
)](q)dq\right)f_0=g_0.$$ Comparing this with Eq. (\ref{e3.1}) we
obtain:
\begin{equation}\label{e3.7a}
    m_0(z;\emptyset )=\alpha ^2 \sigma _1\int {\bar \lambda }(q)
[B(z)^{-1}\sigma _1\lambda (\cdot )](q)dq,
\end{equation}
what provides the analytic continuation of $m(z)$ to $D_{1,\eta
}\setminus I$. We have thus shown that Eq. (\ref{e3.2}) determines
completely the discrete spectrum of the operator $H$ below its
continuous spectrum.

Let us remark that $\lim\limits _{\xi \nearrow e_0+\kappa }m(\xi )$
is finite in dimension $d\geq 3$, because the $1/q^2$-singularity of
the integrand in Eq.(\ref{e3.5}) is integrable. Therefore, in the
case $\varepsilon
>e_0+\kappa $, the second solution $e_1$ of Eq.(\ref{e3.2}) exists
if, and only if, the l.h.s. of that equation is negative for $\xi
=e_0+\kappa $. This exhausts the discrete spectrum of $H$ below its
continuous spectrum.

%%%%%%%%%%%%%%%%%%%%%%%%%%%%%%%%%%%%%%%%%%% Section 4 %%%%%%%%%%%%%%%%%%%%%%%%%%%%%%%%%%%%%%%%%%%%
\section{One-boson branches}\label{sec:4}
\setcounter {equation}{0}

We proceed now to the construction of the one-particle branches of the
(\textit{continuous}) spectrum of the operator $H$. Consider the family $\{
A(\xi ),\xi \in [e_0+\kappa ,\lambda^0_{0,2}-\eta ] \} $ of
selfadjoint operators, acting in ${\cal H}_{\leq 1}$ according to:
for $F=(f_0,f_1)\in {\cal H}_{\leq 1}$,
\begin{equation}\label{A(xi)}
\begin{array}{lcccl}
  (A(\xi )F)_0 & = &  \varepsilon \sigma _3f_0 & + & \alpha \sigma _1\int {\bar \lambda
    }(k) f_1(k) dk \\ & & & & \\
  (A(\xi )F)_1(q) & = & \alpha \sigma _1\lambda (q)f_0 & + & (\varepsilon \sigma _3
  +\omega (q)-m(\xi -\omega (q)))f_1(q) \\ & & & & \\ & & & & -\int {\hat D}_{1,1}(\xi ;q,\emptyset
  ,k)f_1(k)dk.
\end{array}
\end{equation}

Along with this, consider the family $\{ A_0(\xi ),\xi \in
[e_0+\kappa ,\lambda^0_2-\eta ] \} $ acting in ${\cal H}_1$:
\begin{equation}\label{A_0(xi)}
    (A_0(\xi)f)(q)=(\varepsilon \sigma _3
  +\omega (q)-m(\xi -\omega (q)))f(q),\;\; f\in {\cal H}_1.
\end{equation}
Concerning the latter, let us denote by $e_0(\xi ,q),e_1(\xi ,q)$
the eigenvalues of the matrix $\varepsilon \sigma _3 +\omega
(q)-m(\xi -\omega (q))$. One can easily see that these eigenvalues
are the solutions of the equation
\begin{equation}\label{e3.10}
    1+\frac{m_{11}(\xi -\omega (q))+\frac{1}{2\varepsilon }\det{m(\xi -\omega (q))}}{-\varepsilon +\omega
    (q)-e}+\frac{m_{22}(\xi -\omega (q))-\frac{1}{2\varepsilon }\det{m(\xi -\omega (q))}}{\varepsilon +\omega
    (q)-e}=0
\end{equation}
which is similar to Eq. (\ref{e3.2}). The graph of the l.h.s. as a
function of $e$ looks like the graph of Fig. 1.

\includegraphics{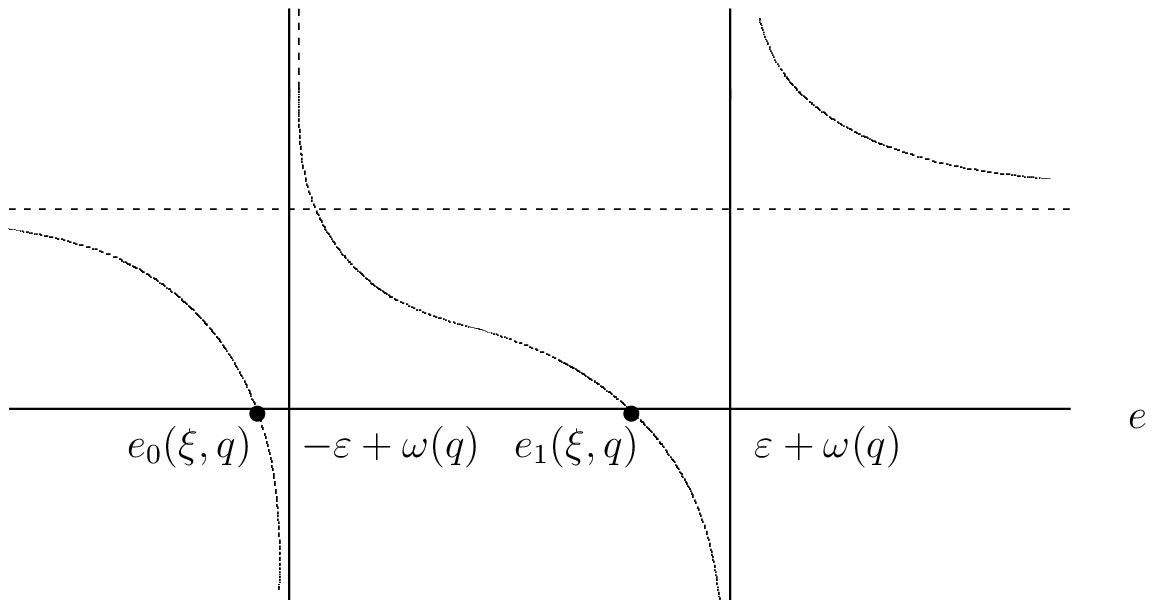}

\begin{center}
\footnotesize{Figure 2: The graph of the l.h.s.\ of (\ref{e3.10})}
\label{fig2}
\end{center}

Hence one can see that its roots are simple and placed in the order
$e_0(\xi ,q)<\omega (q)-\varepsilon <e_1(\xi ,q)<\omega
(q)+\varepsilon $. As, for fixed $q$, the matrix $m(\xi -\omega
(q))$ is positive and increasing of $\xi <\lambda^0_2-\eta $, both
roots $e_0(\xi ,q),e_1(\xi ,q)$ are monotonously decreasing
functions of $\xi $ in the interval $(-\infty ,\lambda^0_{0,2}-\eta
)$. Their graphs are sketched in Fig. 2. Moreover, as a consequence
of Eq. (\ref{e3.10}), the two roots belong respectively to $O(\alpha
^2)$-neighbourhoods of $\omega (q)\mp \varepsilon $, therefore the
distance between them is larger than $\varepsilon $ for small
$\alpha $:
\begin{equation}\label{e3.10a}
    e_1(\xi ,q)-e_0(\xi ,q)>\varepsilon .
\end{equation}
For every $\xi \in [e_0+\kappa ,\lambda^0_{0,2}-\eta )$, let $$E(\xi
)=\inf _q{e_0(\xi ,q)}=-\varepsilon +\kappa -O(\alpha ^2).$$
Obviously, the spectrum of $A_0(\xi )$ is absolutely continuous and
covers the half-axis $[ E(\xi ),\infty ) $.

Let us now calculate the resolvent $(A(\xi )-z)^{-1}$ of $A(\xi )$.
As a preliminary to this, we consider the resolvent of the operator
${\tilde B}(\xi )$ acting in ${\cal H}_1$ as
\begin{equation}\label{e3.11}
    ({\tilde B}(\xi ))(q)=[\varepsilon \sigma _3+\omega (q)
    -m(\xi -\omega (q))]f_1(q)-\int {\hat D}_{1,1}(\xi ;q;\emptyset
    ;k)f_1(k)dk;\; f\in {\cal H}_1.
\end{equation}
In the same way as for Eq. (\ref{e3.5}), we find for $R_{{\tilde
B}(\xi )}(z)=({\tilde B}(\xi )-z)^{-1}$
\begin{equation}\label{e3.12}
\begin{array}{lll}
    (R_{{\tilde B}(\xi )}(z)g)(q) & = & [\varepsilon \sigma _3-z+\omega
    (q)) -m(\xi -\omega (q))]^{-1} \\ & \times & \{ g(q)+\int K_{\xi }(z;q,k)
    [\varepsilon \sigma _3-z+\omega (k)
    -m(\xi -\omega (k))]^{-1}g(k)dk\},
    \end{array}
\end{equation}
where the kernel $K_{\xi }(z;\cdot ,\cdot )\in {\cal K}_1$ is a
${\cal K}_1$-analytic function of $z$ in $\mathbb{C}\setminus [E(\xi
),\infty )$. Thereby, the limits
\begin{equation*}
    K^{\pm }_{\xi }(x;\cdot ,\cdot )=
    \lim\limits _{\epsilon \searrow 0}K_{\xi }(x+i\epsilon ;\cdot ,\cdot )
\end{equation*}
exist in ${\cal K}_1$ and the functions $K^{\pm }_{\xi }(x;\cdot
,\cdot )$ are ${\cal K}_1$-valued H\"{o}lder continuous functions of
$x\in [E(\xi ),\infty )$.

The resolvent $R_{A(\xi )}(z)=(A(\xi )-z\mathbb{I}_{\leq 1})^{-1}$
can now be written as: for $G=(g_0,g_1)$,
\begin{equation}\label{e3.13}
    \begin{array}{cccl}
       (R_{A(\xi )}(z)G)_0 & = & f_0 & =\Delta _{\xi }^{-1}(z)
       (g_0-\alpha \sigma _1\int (R_{{\tilde B}(\xi )}(z)g_1)(k){\bar \lambda }(k)dk)
       \\ & & & \\
       (R_{A(\xi )}(z)G)_1(q) & = & f_1(q) & =(R_{{\tilde B}(\xi )}(z)
       [g_1(\cdot )-\alpha \sigma _1\lambda (\cdot )f_0])(q)
     \end{array}
\end{equation}
where $\Delta _{\xi }(z)$ denotes the ${\cal M}_2$-valued analytic
function of $z\in \mathbb{C}\setminus [E(\xi ),\infty )$
\begin{equation}\label{e3.14}
    \Delta _{\xi }(z)=\varepsilon \sigma _3-z-\alpha ^2\sigma
    _1\int [R_{{\tilde B}(\xi )}(z)\lambda (\cdot )](k){\bar \lambda
    }(k)dk
\end{equation}
The inverse matrix $\Delta _{\xi }^{-1}(z)$ has either one, or two
simple poles $\tau _0(\xi )<\tau _1(\xi )<E(\xi )$ lying on the real
axis at the left of $-\varepsilon $ and $\varepsilon $,
respectively. These poles are the eigenvalues of the operator $A(\xi
)$ with corresponding eigenvectors $\psi _{\xi }^{(0)},\psi _{\xi
}^{(1)}$:
\begin{equation}\label{e3.15}
    \psi _{\xi }^{(i)}=\left( f_{\xi ,0}^{(i)},f_{\xi ,1}^{(i)}(q)=
    -\alpha (R_{{\tilde B}(\xi )}(\tau _i(\xi ))\sigma _1\lambda (\cdot
    ))(q)f_{\xi ,0}^{(i)}\right) ,\; i=0,1,
\end{equation}
where $f_{\xi ,0}^{(i)}$ are the null vectors of the matrix $\Delta
_{\xi }(\tau _i(\xi ))$. As agreed before, we consider the case of
two roots $\tau _0(\xi ),\tau _1(\xi )$. Also, the limits
\begin{equation*}
    [\Delta _{\xi }^{-1}(x)]^{\pm }= \lim\limits _{\epsilon \searrow 0}
    \Delta _{\xi }^{-1}(x\pm i\epsilon ),\; x\in [E(\xi ),\infty )
\end{equation*}
exist, whereby, for $\alpha $ small,
\begin{equation*}
    \pm \textrm{Im}{[\Delta _{\xi }^{\pm }(x)]}=\alpha ^2\pi \sigma _1
    \left( \begin{array}{cc}
    \int\limits _{\omega (k)-\varepsilon  =x}|\lambda (k)|^2 dk& 0 \\
    0 & \int\limits _{\omega (k)+\varepsilon  =x}|\lambda (k)|^2 dk
    \end{array}
    \right)
    +O(\alpha ^4).
\end{equation*}
Hence, as a consequence of assumption (A3), the matrices
${\Delta _{\xi }^{\pm }(x)}^{-1}$ are non-singular on $[E(\xi
),\infty ) $, in other words, the operator $A(\xi )$ has no
eigenvalues embedded in the continuous spectrum.

Remark that, in view of Eq. (\ref{e3.7a}), for $z=\xi $, the matrix
$\Delta _{\xi }(z)$ equals the matrix $\varepsilon  \sigma _3-\xi
-m(\xi )$ in the l.h.s. of Eq. (\ref{e3.1}) and the zeros $\tau
_i(\xi )$ of $\det{\Delta _{\xi }(z)}$ satisfy
\begin{equation}\label{e3.16}
    \tau_i(\xi )=\xi ,\; i=0,1,
\end{equation}
i.e. the condition under which they equal, respectively, the
solutions $e_i,\; i=0,1$ of Eq. (\ref{e3.2}).

Let us construct, for every $\xi \in (e_0+\kappa ,\lambda
^0_{0,2}-\eta )$, the vectors
\begin{equation*}
    F_{\xi }^{(i)}=(\psi _{\xi }^{(i)},{\bar \psi }_{\xi }^{(i)}=
    M_1(\xi )f_{\xi ,1}^{(i)})\in {\cal H},
\end{equation*}
which, obviously, fulfill the equation
\begin{equation*}
    HF_{\xi }^{(i)}=\xi F_{\xi }^{(i)}+(\tau_i(\xi )-\xi )\psi _{\xi
    }^{(i)},
\end{equation*}
therefore, if $\xi =e_i$ they are the eigenvectors of $H$ with
eigenvalues $e_i,\; i=0,1$. In this way, we reobtain the already
constructed eigenvectors of $H$, Eq. (\ref{e3.2a}).
\medskip\

Using general criteria of absence of the \textit{singular continuous}
spectrum of a self-adjoint operator \cite{RS4} and the explicit form
of the resolvent $R_{A(\xi )}(z)$, one can show that $A(\xi )$ has
no singular spectrum. Hence, the space ${\cal H}_{\leq 1}$ splits
into the orthogonal sum of two invariant subspaces of $A(\xi )$:
\begin{equation}\label{e3.17}
    {\cal H}_{\leq 1}={\cal H}^{\rm ac}+{\cal H}^{\rm d},
\end{equation}
where ${\cal H}^{\rm ac}={\cal H}^{\rm ac}(\xi )$ is the subspace of
absolute continuity of $A(\xi )$ and ${\cal H}^{\rm d}={\cal H}^{\rm
d}(\xi )$ is the subspace corresponding to the discrete spectrum,
generated by the vectors $\psi _{\xi }^{(0)},\psi _{\xi }^{(1)}$ (or
by $\psi _{\xi }^{(0)}$ alone in the case of one root).

Consider the embedding
\begin{equation*}
    {\cal I}:{\cal H}_1\rightarrow {\cal H}_{\leq 1}:\;\; {\cal
    I}f=(0,f)\in {\cal H}_{\leq 1} \; {\rm for}\; f\in {\cal H}_1,
\end{equation*}
and construct the wave operator
\begin{equation}\label{e3.18}
    s-\lim\limits _{t\to +\infty}\exp{[itA(\xi )]}{\cal I}\exp{[-itA_0(\xi )]}
    =\Omega^{+}=\Omega^{+}(\xi ).
\end{equation}
The limit exists and $\Omega^{+}:{\cal H}_1\rightarrow {\cal
H}_{\leq 1}$ is unitary and
\begin{equation}\label{e3.19}
    \Omega^{+}A_0(\xi )(\Omega^{+})^{-1}=A(\xi )|_{{\cal H}^{\rm
    ac}}
\end{equation}
i.e. the restriction of $A(\xi )$ to ${\cal H}^{\rm ac}$ is
unitarily equivalent to $A_0(\xi )$ (for details, see \cite{AMZ} or
\cite{Minl2}).

The \textit{generalized} eigenfunctions of the operator $A_0(\xi )$ have the
form
\begin{equation*}
    \delta ^{(i)}_{{\bar q},\xi }=\delta (q-{\bar q})\phi _i(\xi ,{\bar
    q}),
\end{equation*}
where $\phi _i(\xi ,{\bar q})$ are the eigenvectors of the matrix
$\varepsilon  \sigma _3+\omega ({\bar q})-m(\xi -\omega ({\bar q}))$
corresponding to the eigenvalues $e_i(\xi ,{\bar q}),\; i=0,1$ - the
solutions of the equation
\begin{equation}\label{e3.19a}
    \det{[\varepsilon  \sigma _3-z+\omega ({\bar q})-m(\xi -\omega ({\bar
    q}))]}=0.
\end{equation}
Known formulas of scattering theory (\cite{RS3}) allow to write the
generalized eigenfunctions of the continuous spectrum of $A(\xi )$
as
\begin{equation*}
    \psi ^{(i)}_{{\bar q},\xi }=\Omega^{+}\delta ^{(i)}_{{\bar q},\xi }
    =\lim\limits _{\epsilon \searrow 0}i\epsilon
    R_{A(\xi )}(e_i(\xi ,{\bar q})-i\epsilon ){\cal I}\delta ^{(i)}_{{\bar q},\xi
    }.
\end{equation*}
Using the explicit form (\ref{e3.13}) of the resolvent $R_{A(\xi
)}(z)$, we find
\begin{equation}\label{e3.20}
    \begin{array}{ccr}
       \psi ^{(i)}_{{\bar q},\xi ,0} & = & -\alpha \Delta _{\xi }^{-1}(e_i(\xi ,{\bar
       q}))
       \sigma _1 \left[ {\bar \lambda }({\bar q})+\int (\varepsilon  \sigma _3-e_i(\xi ,{\bar q})
       +\omega (k)
        -m(\xi -\omega (k))+i0)^{-1}\right. \\
        &  & \left. \times
        K^-(e_i(\xi ,{\bar q});k,{\bar q}){\bar \lambda (k)dk}\right]
        \phi _i(\xi ,{\bar q})
    \\
    \psi ^{(i)}_{{\bar q},\xi ,1}(q) & = & \delta (q-{\bar q})\phi _i(\xi ,{\bar
    q})+(\varepsilon  \sigma _3-e_i(\xi ,{\bar q})
       +\omega (q)
        -m(\xi -\omega (q)+i0))^{-1}\hfill  \\ & &
        \times K^-(e_i(\xi ,{\bar q});q,{\bar q})
        \phi _i(\xi ,{\bar q})
     \end{array}
\end{equation}

The somewhat formal derivation of the form of the generalized
functions $\psi ^{(i)}_{{\bar q},\xi }$ is bolstered by the
following lemma:

\begin{lemma}\label{L3.1}\hfill
\begin{description}
  \item[(a)] For every fixed ${\bar q}\in \mathbb{R}^d$, $\xi <\lambda ^0_{0,2}-\eta $ and
  $i=0,1$:
  \begin{description}
      \item[i.] the vector $\psi ^{(i)}_{{\bar q},\xi ,0}\in
      \mathbb{C}^2$ is a $\mathbb{C}^2$-valued bounded function of ${\bar
      q}$;
      \item[(ii.)] the function $\psi ^{(i)}_{{\bar q},\xi ,1}(q)$ is a
      $\mathbb{C}^2$-valued generalized function of
      $q$ in ${\tilde {\tilde {{\cal B}'}}}_1$ and, for every fixed $q\in
      \mathbb{R}^d$, it is a $\mathbb{C}^2$-valued generalized function of
      $\bar q$ in ${\tilde {\tilde {{\cal B}'}}}_1$.
    \end{description}
  \item[(b)] For any $\varphi \in {\cal S}(\mathbb{R}^d)$, define
\begin{equation*}
  C_{\varphi ,\xi ,0}^{(i)}=\int \varphi (\bar q)\psi ^{(i)}_{{\bar q},\xi ,0}d{\bar q},\;\;
  C_{\varphi ,\xi ,1}^{(i)}(q)=\int \varphi (\bar q)\psi ^{(i)}_{{\bar q},\xi ,1}(q)d{\bar
  q};
\end{equation*}
then,
\begin{equation}\label{e3.22}
  \Psi _{\varphi ,\xi }^{(i)}=(C_{\varphi ,\xi ,0}^{(i)},C_{\varphi ,\xi ,1}^{(i)}(\cdot
  ))\in {\cal H}_{\leq 1}.
\end{equation}
Thereby, for two functions $\varphi _1,\varphi _2\in {\cal S}(\mathbb{R}^d)$,
\begin{equation}\label{e3.23}
\begin{array}{ccl}
  (\Psi _{\varphi _1,\xi }^{(i)},\Psi _{\varphi _2,\xi }^{(i')}) & = &
  (C_{\varphi _1,\xi ,0}^{(i)},C_{\varphi _2,\xi ,0}^{(i')})_{\mathbb{C}^2}
  +\int (C_{\varphi _1,\xi ,1}^{(i)}(q),C_{\varphi _2,\xi ,1}^{(i')}(q))_{\mathbb{C}^2}dq \\
  & & \\
   & = & (\varphi _1,\varphi _2)_{L_2(\mathbb{R}^d)}\delta _{i,i'}.
\end{array}
\end{equation}
\end{description}
\end{lemma}

A similar lemma appears in \cite{AMZ} and the proof there applies in
our case. The meaning of relation (\ref{e3.23}) can be better seen
by writing it more formally:
\begin{equation}\label{e3.24}
  (\Psi _{{\bar q},\xi }^{(i)},\Psi _{{\bar q'},\xi }^{(i')})
  =\delta ({\bar q}-{\bar q'})\delta _{i,i'}.
\end{equation}
In this way, for every $\xi <\lambda ^0_{0,2}-\eta $ and ${\bar
q}\in \mathbb{R}^d$, we constructed two generalized eigenvectors
$\Psi _{{\bar q},\xi }^{(i)}\in {\tilde {\tilde {{\cal B}'}}}_{\leq
1},\; i=0,1$ of the operator $A(\xi )$ with the eigenvalues
$e_i({\bar q},\xi ),\; i=0,1$, respectively. By acting on them with
the operator $M_1(\xi )$ (see Remark \ref{new}) we obtain the
generalized functions
\begin{equation}\label{e3.25}
  {\bar \Psi }_{{\bar q},\xi }^{(i)}=M_1(\xi )\Psi _{{\bar q},\xi }^{(i)}\in {\tilde
{\tilde {{\cal B}'}}}_{>1}.
\end{equation}
Obviously, the entire vector $F_{{\bar q},\xi }^{(i)}=\{ \Psi _{{\bar q},\xi
}^{(i)},{\bar \Psi }_{{\bar q},\xi }^{(i)}\in {\tilde {\tilde {{\cal B}'}}}\}$
satisfies the equation
\begin{equation*}
  HF_{{\bar q},\xi }^{(i)}=\xi F_{{\bar q},\xi }^{(i)}
  +(e_i({\bar q},\xi )-\xi )\Psi _{{\bar q},\xi }^{(i)}
\end{equation*}
(here $H$ denotes the extension of $H$ to ${\tilde {\tilde {{\cal B}'}}}$, cf. Remark
\ref{new}). Let $\xi ^{(i)}({\bar q}),\; i=0,1$ denote the unique (as seen on Fig. 3)
solutions less than $\pm \varepsilon $ respectively, of the equations
\begin{equation}\label{e3.26}
  e_i({\bar q},\xi )=\xi ,\;\; i=0,1.
\end{equation}

  \includegraphics{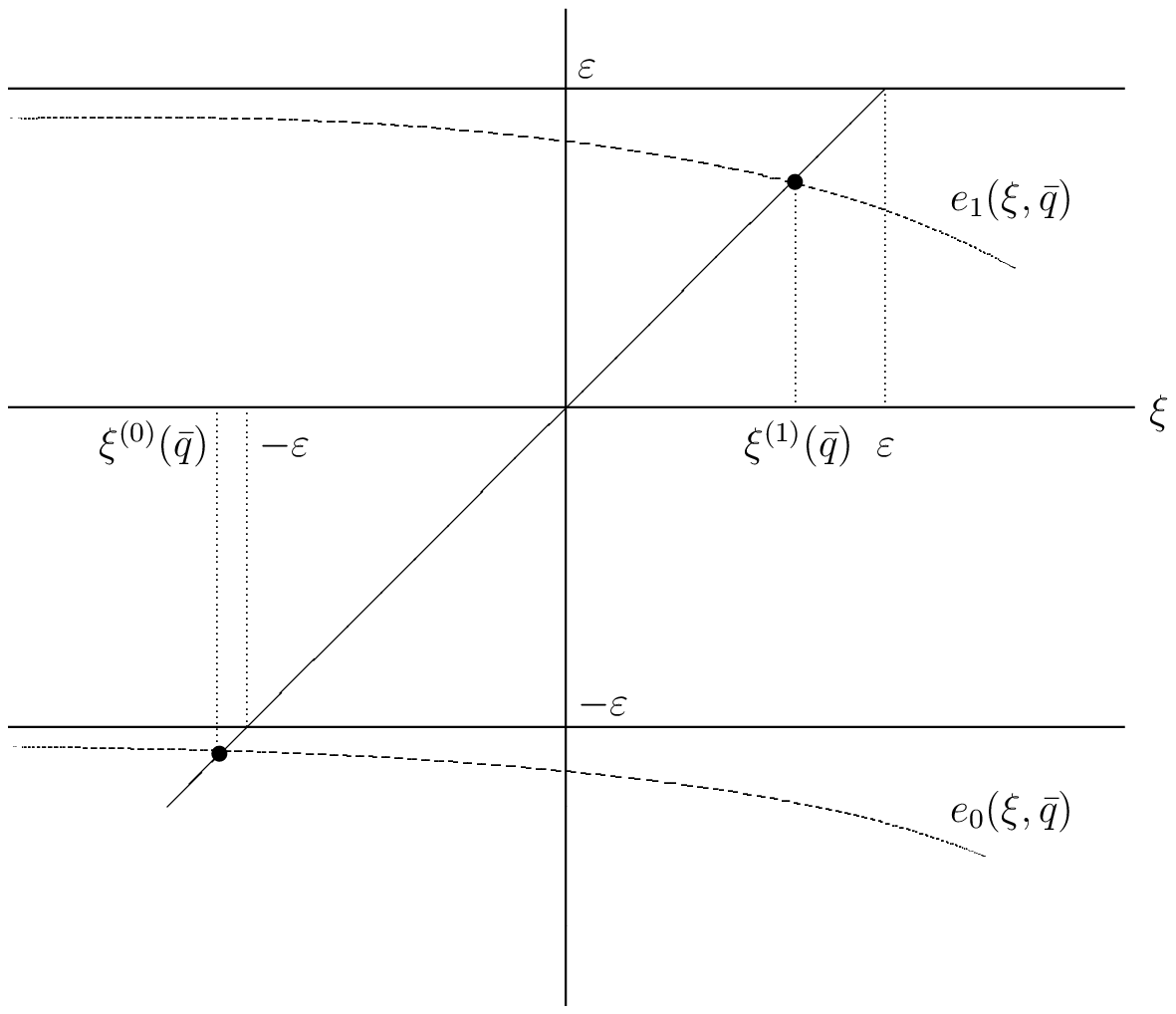}

\begin{center}
\footnotesize{Figure 3: Solving Eq. (\ref{e3.26})}
\label{fig3}
\end{center}
%
% \end{figure}
Now, it is clear that $F_{{\bar q},\xi =\xi ^{(i)}({\bar
q})}^{(i)}=F_{\bar q}^{(i)}$ is a generalized eigenvector of $H$
with eigenvalue $\xi ^{(i)}({\bar q})$. We remind that $e_i({\bar
q},\xi )$ satisfies Eq.(\ref{e3.19a}), which, for $z=\xi $, becomes
\begin{equation}\label{e3.27}
    \det{[\varepsilon \sigma _3-(\xi -\omega (q))-m(\xi -\omega
    (q))]}=0.
\end{equation}
Comparing this equation with Eq. (\ref{e3.1a}), we see that Eq.
(\ref{e3.27}) has two solutions:
\begin{equation}\label{e3.28}
    \xi ^{(i)}({\bar q})=e_i+\omega (q),
\end{equation}
where $e_i,\; i=0,1$ are the two solutions of Eq.(\ref{e3.1a}). Thereby, as the matrix
$\varepsilon \sigma _3+\omega ({\bar q})-m(\xi -\omega ({\bar q}))$ approaches, when
$\xi \rightarrow \xi ^{(i)}({\bar q})$ the matrix $\varepsilon \sigma _3+\omega
(q)-m(e_i)$, one of its eigenvectors $\phi _i(\xi ,{\bar q})$ approaches the
eigenvector of the latter for the eigenvalue $\xi ^{(i)}({\bar q})$, i.e. for all
${\bar q}$ and $i=0,1$, $\;\; \varphi _i(\xi ,{\bar q})|_{\xi =\xi ^{(i)}({\bar
q})}=f_i$.

The conditions
\begin{equation*}
    \xi ^{(i)}({\bar q})<\lambda ^0_{0,2}-\eta ,\;\;\;
\xi ^{(0)}({\bar q})< \xi ^{(1)}({\bar q})
\end{equation*}
define two bounded domains $G_\eta ^{(i)}\subset \mathbb{R}^d$ of
allowed values of $\bar q$:
\begin{equation}\label{e3.29}
    \begin{array}{cc}
       G_\eta ^{(0)}= & \{ {\bar q}:\; \xi ^{(0)}({\bar q})<\lambda ^0_{0,2}-\eta \} \\
       G_\eta ^{(1)}= & \{ {\bar q}:\; \xi ^{(1)}({\bar q})<\lambda ^0_{0,2}-\eta
       \}\subset G_\eta ^{(0)}.
     \end{array}
\end{equation}
For ${\bar q}\in G_\eta ^{(1)}$, there are two generalized
eigenvectors of $H$, $F_{\bar q}^{(i)}=F_{\bar q,\xi ^{(i)}({\bar
q})}^{(i)},\; i=0,1,$ with eigenvalues $\xi ^{(i)}({\bar q}),\;
i=0,1$,respectively, while, for ${\bar q}\in G_\eta ^{(0)}\setminus
G_\eta ^{(1)}$, there is only one eigenvector left. The functions
$\xi ^{(i)}({\bar q})$ are smooth functions of ${\bar q}\in G_\eta
^{(i)}$ and $\min _{\bar q}{\xi ^{(0)}({\bar q})}=e_0+\kappa $
coincides with the lowest end of the continuous spectrum of $H$.

Let ${\cal C}^0_\infty (G_\eta ^{(i)})=:{\cal C}^{(i)},\; i=0,1$ be
the space of infinitely smooth functions with support in the domains
$G_\eta ^{(i)},\; i=0,1$, respectively.
\begin{lemma}
For any $\varphi \in {\cal C}^{(i)}$, the vector
\begin{equation}\label{e3.30}
    F_\varphi ^{(i)}=\int\limits _{G_\eta ^{(i)}} F_{\bar
    q}^{(i)}\varphi ({\bar q}) d{\bar q}\in {\cal H}.
\end{equation}
\end{lemma}
A similar statement is contained in \cite{AMZ} and the proof there
applies to the case at hand through verbatim.

We introduce now two subspaces of ${\cal H}$ as the closures of the
linear span of the vectors (\ref{e3.30}):
\begin{equation*}
    {\cal H}^{(i)}=\overline{\{ F_\varphi ^{(i)},\; \varphi \in {\cal
    C}^{(i)}\}},\;\; i=0,1.
\end{equation*}
As explained in \cite{AMZ}, the calculation of the scalar product of
two vectors (\ref{e3.30}) reduces to the calculation of the
generalized function
\begin{equation}\label{e3.31}
    Q^{i,i'}({\bar q},{\bar q}')=(F_{\bar
    q}^{(i)},F_{{\bar q}'}^{(i')})_{\cal H},\; {\bar q}\in G_\eta ^{(i)},\; {\bar q}'\in
G_\eta
    ^{(i')}.
\end{equation}
As $F_{{\bar q}}^{(i)},F_{{\bar q}'}^{(i')}$ are generalized
eigenvectors of $H$ with eigenvalues $\xi ^{(i)}({\bar q}),\xi
^{(i')}({\bar q}')$ respectively, $Q^{i,i'}({\bar q},{\bar q}')$ is
concentrated on the surface $\{ \xi ^{(i)}({\bar q})=\xi
^{(i')}({\bar q}')\}$, hence it is a sum of generalized functions of
the form $A({\bar q})\delta ({\bar q}-{\bar q}')$ and $B({\bar
q},{\bar q}'))\delta (\xi ^{(i)}({\bar q})-\xi ^{(i')}({\bar q}'))$
(this is rigorously proved in \cite{AMZ}). Therefore, in calculating
$Q^{i,i'}({\bar q},{\bar q}')$, one can discard all terms not
containing such singularities. So,
\begin{equation}\label{e3.32}
\begin{array}{ccl}
  Q^{i,i'}({\bar q},{\bar q}') & = & \left( \psi ^{(i)}_{{\bar q},0},\psi ^{(i')}_{{\bar
    q}',0}\right) _{\mathbb{C}^2}+ \int \left( \psi ^{(i)}_{{\bar q},1}(q),\psi ^{(i')}_{{\bar
    q}',1}(q)\right) _{\mathbb{C}^2}dq \\
   &  &  \\
   & + & \int\limits _{{\cal C}_{>1}}
   \left([M_1(\xi ^{(i)}({\bar q}))\psi ^{(i)}_{{\bar q},1}(\cdot )])(Q),
   [M_1(\xi ^{(i')}({\bar q}'))\psi ^{(i')}_{{\bar q}',1}(\cdot )])(Q)\right)
   _{\mathbb{C}^2}dQ
\end{array}
\end{equation}
(we remind that ${\cal C}_{>1}$ denotes the set of all finite
subsets with more than one point endowed with the measure
(\ref{e1.2})). It will be convenient to calculate separately the
scalar products for each pair $(i,i')=(0,0),(0,1),(1,1)$:
\begin{itemize}
  \item $i=i'=0$. The first term in (\ref{e3.32}) is a continuous
  function of ${\bar q},{\bar q}'$ and will be discarded, as agreed.
  To calculate the second term in (\ref{e3.32}), we represent $\psi ^{(0)}_{{\bar
  q},1}$ as (see (\ref{e3.20}))
  \begin{equation}\label{e3.33}
  \begin{array}{ccc}
  \psi ^{(0)}_{{\bar q},1}(q) & = & \delta (q-{\bar q})f_0+
  [A_0(\xi ^{(0)}({\bar q}),q,{\bar q})f_0]\diagup [\xi ^{(0)}(q)-\xi ^{(0)}({\bar q})+i0] \\
   &  &  \\
   &  & +[A_1(\xi ^{(0)}({\bar q}),q,{\bar q})\varphi _1^0(\xi ^{(0)}({\bar
   q}),q)]\diagup [e_1(\xi ^{(0)}({\bar q}),q)-\xi ^{(0)}({\bar q})+i0]
  \end{array}
  \end{equation}
and similarly $\psi ^{(0)}_{{\bar q}',1}$; here, $f_0$ is the null
vector of the matrix (\ref{e3.1}) corresponding to the solution
$e_0$ of Eq. (\ref{e3.2}), $e_1(\xi ^{(0)}({\bar q}),q)$ is the
second eigenvalue of the matrix $\varepsilon \sigma _3 +\omega
(q)-m(\xi ^{(0)}({\bar q})-\omega (q))$ and $\varphi _1^0(\xi
^{(0)}({\bar q}),q)$ the eigenvector corresponding to it, and
\begin{equation}\label{e3.34}
    \begin{array}{ccl}
       A_0(\xi ^{(0)}({\bar q}),q,{\bar q}) & = &
       (K^-(\xi ^{(0)}({\bar q}),q,{\bar q})f_0,f_0)_{\mathbb{C}^2} \\
        &  &  \\
       A_1(\xi ^{(0)}({\bar q}),q,{\bar q}) & = & (K^-(\xi ^{(0)}({\bar q}),q,{\bar q})
       \varphi _1(\xi ^{(0)}({\bar q}),q),\varphi _1(\xi ^{(0)}({\bar
       q}),q)_{\mathbb{C}^2}.
     \end{array}
\end{equation}
By virtue of Eq. (\ref{e3.10a}), the third term in (\ref{e3.33}) is
a regular function of $q$ and $\bar q$ and we discard it. For the
second term we use Sokhotski's formula $1/[x+i0]=i\pi \delta (x)
+P(1/x)$ and obtain
\begin{equation*}
    \psi ^{(0)}_{{\bar q},1}(q)=\delta (q-{\bar q})f_0+i\pi A_0(\xi ^{(0)}({\bar q}),q,{\bar q})
    \delta (\xi ^{(0)}({q})-\xi ^{(0)}({\bar q}))f_0+{\rm
    regular}\;{\rm
    terms}.
\end{equation*}
In this way, the second term in Eq. (\ref{e3.32}) equals, modulo
regular terms:
\begin{equation}\label{e3.35}
\begin{array}{l}
  \int \left( \psi ^{(0)}_{{\bar q},1}(q),\psi ^{(0)}_{{\bar
    q}',1}(q)\right) _{\mathbb{C}^2}dq = \delta ({\bar q}-{\bar q}')
    \\
    \\ \hfill +\delta (\xi ^{(0)}({\bar q}')-\xi ^{(0)}({\bar q}))
    \left\{ i\pi [A_0(\xi ^{(0)}({\bar q}),{\bar q}',{\bar q})
    -\overline{A_0(\xi ^{(0)}({\bar q}'),{\bar q},{\bar
    q}')}]\right.
     \\ \\ \left.
   + \pi ^2\int\limits _{\xi ^{(0)}({q})=\xi ^{(0)}({\bar q})} A_0(\xi ^{(0)}({\bar q}),q,{\bar q})
    \overline{A_0(\xi ^{(0)}({\bar q}'),q,{\bar q}')
    }dq \right\}
\end{array}
\end{equation}
For the calculation of the third term of Eq. (\ref{e3.32}) we
represent $(M_1(\xi ^{(0)}({\bar q}))\psi ^{(0)}_{{\bar q},1}(\cdot
))(Q) $ as (see Eq. (\ref{e1.10})):
\begin{equation}\label{e3.36}
\begin{array}{ccc}
  (M_1(\xi ^{(0)}({\bar q}))\psi ^{(0)}_{{\bar q},1}(\cdot
))(Q) & = & \sum\limits _{q\in Q}\mu ^{(0)}_1(\xi ^{(0)}({\bar
q});Q\setminus q;q;\emptyset )\psi ^{(0)}_{{\bar q},1}(q) \\
   & + & \int \mu ^{(1)}_1(\xi ^{(0)}({\bar
q});Q;\emptyset ;k)\psi ^{(0)}_{{\bar q},1}(k)dk.
\end{array}
\end{equation}
Now, it is easy to see that the only contribution of this expression
to the third term of (\ref{e3.32}) comes from the sum over $q$ and
equals
\begin{equation}\label{e3.37}
    \begin{array}{r}
    \int\limits _{\mathbb{R}^d}dq\int\limits _{{\cal C}_{>1}}dQ
    (\mu ^{(0)}_0(\xi ^{(0)}({\bar q})-\omega (q);Q;\emptyset ;\emptyset )f_0,
    \mu ^{(0)}_0(\xi ^{(0)}({\bar q}')-\omega (q);Q;\emptyset ;\emptyset )f_0)
    _{\mathbb{C}^2} \\
    \times [\delta (q-{\bar q})+i\pi A_0(\xi ^{(0)}({\bar q});q,{\bar q})
    \delta (\xi ^{(0)}({q})-\xi ^{(0)}({\bar
    q}))] \\ \times [\delta (q-{\bar q}')+i\pi A_0(\xi ^{(0)}({\bar q}');q,{\bar q}')
    \delta (\xi ^{(0)}({q})-\xi ^{(0)}({\bar
    q}))] \\ =\int dQ\left( \mu ^{(0)}_0(e_0;Q;\emptyset ;\emptyset )f_0,
    \mu ^{(0)}_0(e_0;Q;\emptyset ;\emptyset )f_0\right)
    _{\mathbb{C}^2}\hfill
    \\
    \times \left\{ \delta ({\bar q}-{\bar q}')\right. +
    \left[ i\pi (A_0(\xi ^{(0)}({\bar q});{\bar q}',{\bar q})
    -\overline{A_0(\xi ^{(0)}({\bar q}');{\bar q},{\bar
    q}')})\right. \hfill
    \\ \left. \left.
    +\pi ^2 \int_{\{ \xi ^{(0)}({q})=\xi ^{(0)}({\bar q})\} }
    dq A_0(\xi ^{(0)}({\bar q});q,{\bar q})A_0(\xi ^{(0)}({\bar q}');q,{\bar q}')
     \right] \delta (\xi ^{(0)}({\bar q})-\xi ^{(0)}({\bar q}'))\right\}
    \end{array}
\end{equation}
In the equality above we used Eq. (\ref{mun_mum}). Let us remark
that the vector
\begin{equation*}
    F_0(Q)=\left\{ \begin{array}{rc}
                     f_0, & Q=\emptyset  \\
                     \mu ^{(0)}_0(e_0,Q,\emptyset ,\emptyset )f_0, &
                     |Q|>0
                   \end{array}
    \right.
\end{equation*}
is nothing but the eigenvector (found above) of $H$ corresponding to
$e_0$. Hence,
\begin{equation*}
\begin{array}{r}
    \int dQ\left( \mu ^{(0)}_0(e_0;Q;\emptyset ;\emptyset )f_0,
    \mu ^{(0)}_0(e_0;Q;\emptyset ;\emptyset )f_0\right)
    _{\mathbb{C}^2}\hfill \\ \\ =\| F_0\| ^2-\| f_0\| ^2=\| F_0\| ^2-1=R=O(\alpha
    ^2).
    \end{array}
\end{equation*}
Collecting the expressions above, we obtain that $Q^{0,0}({\bar
q},{\bar q}')$ equals
\begin{equation*}
    \begin{array}{crl}
       Q^{0,0}({\bar q},{\bar q}') & =\hfill \| F_0\| ^2 &
       \left[ \delta ({\bar q}-{\bar q}')\right. \\
        &   & +\left\{ i\pi (A_0(\xi ^{(0)}({\bar q});{\bar q}',{\bar q})-
    \overline{A_0(\xi ^{(0)}({\bar q}');{\bar q},{\bar q}')})\right. \\
       &   &
       \left. +\pi ^2\int _{\{ \xi ^{(0)}({q})=\xi ^{(0)}({\bar q})\} }
       dq A_0(\xi ^{(0)}({\bar q});q,{\bar q})\overline{
       A_0(\xi ^{(0)}({\bar q}');q,{\bar q}')}
     \right\} \\ & & \\ & & \left. \times \delta (\xi ^{(0)}({\bar q}')-\xi ^{(0)}
     ({\bar q}))\right]
     \end{array}
\end{equation*}

  \item $i=i'=1$. A similar calculation gives
  \begin{equation*}
    \begin{array}{crl}
       Q^{1,1}({\bar q},{\bar q}') & =\hfill \| F_1\| ^2 &
       \left[ \delta ({\bar q}-{\bar q}')\right. \\
        &   & +\left\{ i\pi (A_1(\xi ^{(1)}({\bar q});{\bar q}',{\bar q})-
    \overline{A_1(\xi ^{(1)}({\bar q}');{\bar q},{\bar q}')})\right. \\
       &   &
       \left. +\pi ^2\int _{\{ \xi ^{(1)}({q})=\xi ^{(1)}({\bar q})\} }
       dq A_1(\xi ^{(1)}({\bar q});q,{\bar q})\overline{
       A_1(\xi ^{(1)}({\bar q}');q,{\bar q}')}
     \right\} \\ & & \\ & & \left. \times \delta (\xi ^{(1)}({\bar q}')-\xi ^{(1)}
     ({\bar q}))\right],
     \end{array}
\end{equation*}
where $F_0$ and $F_1$ are the eigenvectors of $H$ constructed above,
corresponding to the eigenvalues $e_0$ and $e_1$, respectively.
  \item $i\neq i'$. Finally,
  \begin{equation}\label{e3.38}
    Q^{0,1 }({\bar q},{\bar q}')=Q^{1,0}({\bar q}',{\bar q})=0,
  \end{equation}
  as these functions have as factors $(F_0,F_1)_{\cal H}=0$.
\end{itemize}
We have thus proved:
\begin{lemma}\label{L3.2}
The subspaces ${\cal H}^{(1)}$ and ${\cal H}^{(2)}$ are orthogonal.
The scalar product in each of them has the form
\begin{equation}\label{e3.39}
\begin{array}{clc}
  (F^{(i)}_{\varphi _1},F^{(i)}_{\varphi _2})_{{\cal H}^{(i)}} &
   =\| F_i\| ^2\int\limits _{\kappa +e_i}^{\lambda ^0_{0,2}-\eta }
   dx\left[ \int _{\chi _x^{(i)}}\varphi _1(q)
   \overline {\varphi _2(q)}d\nu _x^{(i)}(q)\right. &  \\ & & \\
   & \left. + \int _{\chi _x^{(i)}}\int _{\chi _x^{(i)}}M_x^{(i)}(q,q')
   \varphi _1(q)\overline {\varphi _2(q')}d\nu _x^{(i)}(q)d\nu _x^{(i)}(q')\right] ,
   & i=0,1.
\end{array}
\end{equation}
Here, $\chi _x^{(i)}$ is the level surface of the function $\xi ^{(i)}(\cdot )$:
\begin{equation*}
  \chi _x^{(i)}=\{ q:\; \xi ^{(i)}(q)=x\} ,\;\; x\in [x+e_1,\lambda ^0_{0,2}-\eta ),
\end{equation*}
endowed with the Gelfand-Leray measure $d\nu _x^{(i)}$ generated by the function $\xi
^{(i)}(\cdot )$, and $M_x^{(i)}({\bar q},{\bar q}')$ denotes the restriction to $\chi
_x^{(i)}\times \chi _x^{(i)}$ of the function
\begin{equation}\label{e3.40}
    \begin{array}{ccl}
  M^{(i)}({\bar q},{\bar q}') & = & i\pi (A_i(\xi ^{(i)}({\bar q});{\bar q}',{\bar q})-
    \overline{A_i(\xi ^{(i)}({\bar q}');{\bar q},{\bar q}')}) \\ & + &
        \pi ^2\int _{\{ \xi ^{(i)}({q})=\xi ^{(i)}({\bar q})\} }
       dq A_i(\xi ^{(i)}({\bar q});q,{\bar q})\overline{
       A_i(\xi ^{(i)}({\bar q}');q,{\bar q}')}.
    \end{array}
\end{equation}
\end{lemma}

Our estimates imply that the operators $M^{(i)},\; i=0,1$ given by the kernels
$M^{(i)}({\bar q},{\bar q}'),\; i=0,1$ are bounded in ${\cal H}^{(i)},\; i=0,1$,
respectively, and their norms are $\leq 1$ for small $\alpha $. This, and the formulas
(\ref{e3.38}), (\ref{e3.39}), imply, in particular, that
\begin{equation*}
  C_0\|\varphi \| _{L_2(G^{(i)}_{\eta })}<\| F^{(i)}_{\varphi }\| _{{\cal H}^{(i)}}<
  C_1\|\varphi \| _{L_2(G^{(i)}_{\eta })}
\end{equation*}
for certain constants $0<C_0<C_1$, therefore the applications
$\varphi \mapsto F^{(i)}_{\varphi }:{\cal C}^{(i)}\rightarrow {\cal
H}^{(i)}$ are continuous with respect to the $L_2(G^{(i)}_{\eta
})$-norm, and, as such, they extend to one-to-one applications
$:L_2(G^{(i)}_{\eta })\rightarrow {\cal H}^{(i)}$, $\varphi \mapsto
F^{(i)}_{\varphi },\; \varphi \in L_2(G^{(i)}_{\eta })$. Thereby,
the action of the operator $H$ on a vector $F^{(i)}_{\varphi }\in
{\cal H}^{(i)}$ is given by the formula
\begin{equation}\label{e3.41}
  HF^{(i)}_{\varphi }=F^{(i)}_{\xi ^{(i)}\varphi },
\end{equation}
where $(\xi ^{(i)}\varphi )(q)=\xi ^{(i)}(q)\varphi (q),\; q\in G^{(i)}_{\eta }$.

As seen from Eq. (\ref{e3.39}), each of the spaces ${\cal H}^{(i)}$
can be represented as a direct integral of spaces
\begin{equation*}
  {\cal H}^{(i)}=\int\limits _{[e_i+\kappa ,\lambda ^0_{0,2}-\eta ]}^{\bigoplus }{\cal
  G}_x^{(i)}dx,\;\; {\rm where}\;\; {\cal
  G}_x^{(i)}=L_2(\chi _x^{(i)},d\nu _x^{(i)}),
\end{equation*}
whereby the spaces ${\cal G}_x^{(i)}$ are "eigenspaces" of $H^{(i)}=H|_{{\cal
H}^{(i)}}$, i.e.
\begin{equation*}
  H^{(i)}=\int\limits _{[e_i+\kappa ,\lambda ^0_{0,2}-\eta ]}^{\bigoplus }xI_x^{(i)}dx,
\end{equation*}
where $I_x^{(i)}$ is the unit operator in ${\cal G}_x^{(i)}$.

\begin{lemma}\label{L3.3}
For each $i=0,1$ there exists a bounded operator $B^{(i)}$ acting in ${\cal H}^{(i)}$
and commuting with $H^{(i)}$, such that
\begin{equation}\label{e3.42}
  \left( F^{(i)}_{B^{(i)}\varphi _1},F^{(i)}_{B^{(i)}\varphi _2}\right)=
  \left( \varphi _1,\varphi _2\right)_{L_2(G^{(i)}_{\eta })} ,\;\; \varphi _1,\varphi _ 2\in L_2(G^{(i)}_{\eta })
\end{equation}
\end{lemma}
\textit{Proof.} The scalar product (\ref{e3.39}) induces in ${\cal G}_x^{(i)}$ a
sesquilinear form:
\begin{equation*}
  \langle \varphi _1,\varphi _2\rangle _{{\cal G}_x^{(i)}}
  =\| F^{(i)}\| _2((I_x^{(i)}+M_x^{(i)})\varphi _1,\varphi _2)_{{\cal
  G}_x^{(i)}},
\end{equation*}
where $(\varphi _1,\varphi _2)_{{\cal G}_x^{(i)}}$ is the scalar
product in ${\cal G}_x^{(i)}$ and $M_x^{(i)}$ is the selfadjoint
operator in ${\cal G}_x^{(i)}$ defined by the kernel
$M^{(i)}_x({\bar q},{\bar q}')$. Our estimates show that $\|
M_x^{(i)}\| <1$ for $\alpha $ small, hence that the bounded operator
$B_x^{(i)}=(I_x^{(i)}+M_x^{(i)})^{-1/2}$ exists. Obviously,
\begin{equation*}
  \langle B_x^{(i)}\varphi _1,B_x^{(i)}\varphi _2\rangle _{{\cal G}_x^{(i)}}
  =(\varphi _1,\varphi _2)_{{\cal G}_x^{(i)}}.
\end{equation*}
Therefore, the operator $B^{(i)}=\int\limits _{[e_i+\kappa ,\lambda
^0_{0,2}-\eta ]}^{\bigoplus }B_x^{(i)} dx$, which commutes with
$H^{(i)}$, satisfies the condition (\ref{e3.42}) as well. The lemma
is proved.\hfill $\square$

The Lemma \ref{L3.3} shows that the application $\varphi \mapsto {\hat F}_\varphi
:L_2(G^{(i)}_\eta )\rightarrow {\cal H}^{(i)}$, where ${\hat F}_\varphi =
F_{B^{(i)}\varphi },\; \varphi \in L_2(G^{(i)}_\eta )$, is unitary. Thereby, as
$B^{(i)}$ commutes with $H^{(i)}$, the relation (\ref{e3.41}) still holds:
\begin{equation*}
  H{\hat F}^{(i)}_{\varphi }={\hat F}^{(i)}_{\xi ^{(i)}\varphi }.
\end{equation*}

%%%%%%%%%%%%%%%%%%%%%%%%%%%%%%%%%%%%%%%%%%% Section 5 %%%%%%%%%%%%%%%%%%%%%%%%%%%%%%%%%%%%%%%%%%%%
\section{Conclusion}\label{sec:5}
\setcounter {equation}{0}

In conclusion we would like to notice that observations made in Sections \ref{sec:3}
and \ref{sec:4} can be resumed as the following statement:

\begin{theorem}\label{main Th}
Under the assumptions made concerning the parameters of the model (\ref{Ham-sec-quant}),
and for $\alpha $ sufficiently small, for the operator $H$ one has:\\
{\rm{(}i\rm{)}} One (or two) eigenvectors $F_0^{(i)}$ with the eigenvalues $e_i$, respectively,
  below its continuous spectrum {\rm{(}}$i=0$, or $i=0,1${\rm{)}}.\\
{\rm{(}ii\rm{)}} For any $\eta $, one, or two (depending on the number of eigenvectors
  $F_0^{(i)}$), invariant subspaces ${\cal H}^{(i)}$, and one, or two, bounded domains
  $G^{(i)}_\eta \subset \mathbb{R}^d$, such that the restriction of $H$ to ${\cal
  H}^{(i)}$is unitarily equivalent to the operator of multiplication by the function
\begin{equation*}
  \xi ^{(i)}(q)=e_i+\omega (q),\; q\in G^{(i)}_\eta
\end{equation*}
acting in $L_2(G^{(i)}_\eta )$.
\end{theorem}
This theorem summarizes our main result about the structure of the spectrum of a
two-level quantum system weakly coupled to a boson field (spin-boson model)
announced in the Section \ref{sec:1}.

{In conclusion we would like to note that (in spite of the technical difficulties) we believe that our method
allows some generalizations and improvements to be able :\\
(i) to construct in a similar way the multi-boson branches; cf. \cite{DG}, where it is nicely done in a different way
(a limited space of the present paper does not allow us to enter into details of \cite{DG} and
we recommend it for the reader as an important reference); \\
(ii) to prove the completeness, which reduces to the proof that the Hilbert space
${\cal H}=\mathbb{C}^2 \otimes {\cal F}_s $ in imbedded by a nuclear operator
into the space ${\tilde{\tilde {{\cal B}'}}}$, see Remark \ref{new} and (\ref{dual space B}).}

{A project concerning these two points is now in progress. }

%%%%%%%%%%%%%%%%%%%%%%%%%%%%%%%%%%%%%%% Acknowledgements %%%%%%%%%%%%%%%%%%%%%%%%%%%%%%%%%%%%%%%%%%%%%%%%%%%%%
\vskip 1.5cm

\noindent {\bf Acknowledgements.} N. A. and R. A. M. thank Centre de Physique
Th\'{e}orique (UMR 6207) - Luminy, where this work was initiated, for financial support and
the warm hospitality. N.A. acknowledges
the financial support of the CERES and CEEX programs (Grants No. 4-187/2004 and 05-D11-06).
R. A. M. acknowledges the financial support from the funds RFFI and
CRDF.

We would like to thank referees for very relevant remarks and constructive criticism. They motivated us
to take a time to revise and to make more precise the presentation of results that we are able to proof
at the present time.

\newpage
%%%%%%%%%%%%%%%%%%%%%%%%%%%%%%%%%%%%%%%%% Appendix %%%%%%%%%%%%%%%%%%%%%%%%%%%%%%%%%%%%%%%%%%%%%%%%%%%%%%
\section{Appendix}\label{App}
\setcounter {equation}{0}

\textit{Proof of Proposition \ref{P3.1} }:

\vskip 0.2cm

\noindent Let $\mathcal{K}_1$ be a space of smooth functions: $f(q,k) \in \mathcal{M}_2$ for $q,k \in {\mathbb{R}}^d$,
such that
\begin{equation}\label{A1}
\|(\partial_{q}^{\varepsilon_1}\partial_{k}^{\varepsilon_2} f)(q,k)\| \leq C \ h(q)h(k) \ , \  \  \
\varepsilon_{1,2} = 0, 1 \ .
\end{equation}
Here $\partial_{q}^{\varepsilon} : = (\partial_{q_1}^{\varepsilon}, \ldots,  \partial_{q_d}^{\varepsilon})$ and
$\|\cdot\|$ is equal to the sum of the matrix-norms of
$\{\partial_{q_i}^{\varepsilon_1}\partial_{q_j}^{\varepsilon_2} \ f \}_{i,j = 1, \varepsilon_1, \varepsilon_2}^{d}$.
We define the norm in $\mathcal{K}_1$ by
\begin{equation}\label{norm K}
\|f\|_{\mathcal{K}_1} := \inf \ C  \ ,
\end{equation}
over $C$ verifying (\ref{A1}).

Recall that the kernels $\{{\hat D}_{1,1}(z;q;\emptyset ;k)\}_z$ belong to the family of $\mathcal{K}_1$-valued
functions $\mathcal{D}_{z}(q,k)$ defined in the semi-plane
\begin{equation}\label{semi-plane}
D_{1,\eta} = \{z\in \mathbb{C}: \textrm{Re} \,z < \lambda_{0,2}^{0} - \eta = 2 \kappa - \varepsilon - \eta\} \ ,
\end{equation}
see (\ref{Dn,eta}), with the cut along the interval (see Proposition \ref{P3.1}):
\begin{equation}\label{cut-interval}
I = (\kappa + e_0, \ 2 \kappa - \varepsilon - \eta) \ .
\end{equation}
We assume that:\\
(1) The family $\{\mathcal{D}_{z}(q,k)\}_z$ is $\mathcal{K}_1$-analytic in $D_{1,\eta} \backslash I$.\\
(2) It is also $\mathcal{K}_1$-continuous and bounded: $\|\mathcal{D}_{z}\|_{\mathcal{K}_1} < L_1$,
in the closure $\overline{D_{1,\eta} \backslash I}$.\\
(3) $\lim_{z\rightarrow 0} \|\mathcal{D}_{z}\|_{\mathcal{K}_1} = 0$ .\\
(4) For $x \in I$ the limit values $\mathcal{D}_{x}^{\pm}(q,k):=\lim_{z\rightarrow \pm x} \mathcal{D}_{z}(q,k)$
verify the H\"{o}lder condition:
\begin{equation}\label{Hold-on-I}
|\mathcal{D}_{x_1}^{\pm}(\cdot, \cdot) - \mathcal{D}_{x_2}^{\pm}(\cdot, \cdot)|  \ < \ C_1 \ |x_1 - x_2|^{1/2} \ ,
\end{equation}
for constant $C_1$ and exponent $1/2$.

By (\ref{e3.4}) the kernel of the inverse operator $B(z)^{-1}$ has the form:
\begin{eqnarray}\label{inv-Kern}
[B(z)^{-1}](q,k) &=& [\varepsilon \sigma_3-(z-\omega (q))-m(z-\omega (q))]^{-1} \\
&\times & \{\delta(q,k) + K(z;q,k) [\varepsilon \sigma_3-(z-\omega (k))-m(z-\omega (k))]^{-1}\} \ , \nonumber
\end{eqnarray}
where (\textit{formally}) we put:
\begin{equation}\label{K-series}
K(z;q,k)= \sum_{n=1}^{\infty} \ K^{(n)}(z;q,k)
\end{equation}
with
$K^{(n=1)}(z;q,k):= \mathcal{D}_{z}(q,k)$ and the other terms are defined by recursions
\begin{equation}\label{recursion}
K^{(n+1)}(z;q,k):=\int_{{\mathbb{R}}^d} K^{(n)}(z;q,q')
[\varepsilon \sigma_3-(z-\omega (q'))-m(z-\omega (q'))]^{-1} K^{(1)}(z;q',k)\ dq' \ ,
\end{equation}
for $n+1 \geq 2$.

Let for a given $n$ the kernel $K^{(n)}(z;q,q')$ verifies the same conditions
(1)-(4) as the function $\mathcal{D}_{z}(q,k)$, we denote them $(1)^{(n)}-(4)^{(n)}$ with the evident substitutions of the
corresponding constants by $L_n$ and $C_n$. Now we have to prove that the function $K^{(n+1)}(z;q,q')$
satisfies conditions $(1)^{(n+1)}-(4)^{(n+1)}$ with constants that verify the estimates:
\begin{equation}\label{estim}
L_{n+1} < \, \theta \ L_{n} \ \ \ \textrm{ and } \ \ \ C_{n+1} < \, \theta \ C_{n} \ \ ,
\end{equation}
for some $0<\theta<1$. Notice that (\ref{estim}) implies (a uniform in the closure $\overline{D_{1,\eta} \backslash I}$)
convergence of (\ref{K-series}) in the $\mathcal{K}_1$-norm, i.e. the the kernel $K(z;q,q')$ verifies the conditions
(1)-(4) for some $L_{\infty}$ and $C_{\infty}$.

To this end we first use the representation:
\begin{equation}\label{repres1}
[\varepsilon \sigma_3-(z-\omega (q'))-m(z-\omega (q'))]^{-1} = \frac{A_0}{e_0 - z - \omega (q')} +
\frac{A_1}{e_1 - z - \omega (q')} + \varphi (z-\omega (q')) \ .
\end{equation}
Here $A_0$ and $A_1$ are two \textit{residues} of the matrix-valued rational complex function
$\Phi(\zeta):= [\varepsilon \sigma_3-\zeta-m(\zeta)]^{-1}$  at poles $e_0$ and $e_1$ respectively, see (\ref{e3.1a}),
and $\varphi: \mathbb{C} \mapsto \mathcal{M}_2$ is an analytic function bounded in domain
$\{\zeta\in \mathbb{C}: \textrm{Re}\, \zeta < \lambda_{0,1}^{0}-\eta = \kappa - \varepsilon - \eta\}$. Then
(\ref{recursion}) yields:
\begin{equation}\label{recution-0-1}
K^{(n+1)}(z;q,k)= K^{(n+1),0}(z;q,k) + K^{(n+1),1}(z;q,k) + \widehat{K}^{(n+1)}(z;q,k)  \ ,
\end{equation}
where ($j=0,1$)
\begin{eqnarray}\label{K-bis}
&&K^{(n+1),j}(z;q,k) := \int_{{\mathbb{R}}^d} \, K^{(n)}(z;q,q') \
\frac{A_j}{e_j - z - \omega (q')}\ K^{(1)}(z;q',k)\ dq'  \  ,\\
&&\widehat{K}^{(n+1)}(z;q,k) := \int_{{\mathbb{R}}^d} \, K^{(n)}(z;q,q') \
\varphi (z-\omega (q')) \ K^{(1)}(z;q',k)\ dq' \ .
\label{K}
\end{eqnarray}
The recursions (\ref{K-bis}) and (\ref{K}) imply $\mathcal{K}_1$-analyticity of terms (\ref{recution-0-1})
in domain $D_{1,\eta} \backslash I$ and the estimates:
\begin{eqnarray*}
&&\max \{\|\partial_{z} K^{(n+1),j}(z;q,k)\|_{{\mathcal{K}}_1}; \|K^{(n+1),j}(z;q,k)\|_{{\mathcal{K}}_1}\} <
Const \times\\
&&\max \{\|\partial_{z} K^{(n),j}(z;q,k)\|_{{\mathcal{K}}_1}; \|K^{(n),j}(z;q,k)\|_{{\mathcal{K}}_1}\}
\max \{\|\partial_{z} K^{(1),j}(z;q,k)\|_{{\mathcal{K}}_1}; \|K^{(1),j}(z;q,k)\|_{{\mathcal{K}}_1}\} \times \\
&&\left\{\frac{1}{(dist (z, I)) ^2} +  \frac{1}{dist (z, I)}\right\}  \ ,
\end{eqnarray*}
where $dist (z, I)$ is the distance between $z$ and the cut $I$. Similarly one also obtains:
\begin{eqnarray*}
&&\max \{\|\partial_{z} \widehat{K}^{(n+1)}(z;q,k)\|_{{\mathcal{K}}_1};
\|\widehat{K}^{(n+1)}(z;q,k)\|_{{\mathcal{K}}_1}\} <
Const \times\\
&&\max \{\|\partial_{z} \widehat{K}^{(n)}(z;q,k)\|_{{\mathcal{K}}_1};
\|\widehat{K}^{(n)}(z;q,k)\|_{{\mathcal{K}}_1}\}
\max \{\|\partial_{z} \widehat{K}^{(1)}(z;q,k)\|_{{\mathcal{K}}_1};
\|\widehat{K}^{(1)}(z;q,k)\|_{{\mathcal{K}}_1}\} \times \\
&& \sup_{\zeta \in (D_{1,\eta} \backslash I)} |\varphi (\zeta)| \ .
\end{eqnarray*}
Moreover, for any $n$ the function $\widehat{K}^{(n)}(z;q,k)$ is $\mathcal{K}_1$-continuous on the closure
$\overline{D_{1,\eta} \backslash I}$ and one gets for the limit values on the cut:
\begin{eqnarray}\label{Lim-K-cut}
&&\widehat{K}^{(n+1),\pm}(x;q,k):= \lim_{z\rightarrow \pm x} \widehat{K}^{(n+1)}(z;q,k)= \\
&&\int_{{\mathbb{R}}^d} \, K^{(n),\pm}(x;q,q') \ \varphi (x-\omega (q')) \ K^{(1),\pm}(x;q',k)\ dq' \ . \nonumber
\end{eqnarray}
Then estimate (\ref{Hold-on-I}), condition $(4)^{(n)}$ , and (\ref{Lim-K-cut}) yield
\begin{equation}\label{Holder-K-hat}
\|\widehat{K}^{(n+1),\pm}(x_1;\cdot,\cdot) - \widehat{K}^{(n+1),\pm}(x_2;\cdot,\cdot)\|_{{\mathcal{K}}_1}  \ <
\ R_{n+1} \ |x_1 - x_2|^{1/2} \ ,
\end{equation}
where
\begin{eqnarray*}
R_{n+1} &=& Const \ \max\{C_n , \sup_{x \in I} \|{K}^{(n),\pm}(x;\cdot,\cdot)\|_{{\mathcal{K}}_1}\}  \times \\
&&\max\{C_1 , \sup_{x \in I} \|{K}^{(1),\pm}(x;\cdot,\cdot)\|_{{\mathcal{K}}_1}\} \
\max_{\zeta \in (\overline{D_{1,\eta} \backslash I})} |\varphi (\zeta)| \ .
\end{eqnarray*}
By virtue of (\ref{Lim-K-cut}) one also gets that
$\lim_{z\rightarrow\infty}\|\widehat{K}^{(n+1)}(z;\cdot,\cdot)\| = 0 $. Therefore, the family
$\{\widehat{K}^{(n+1)}(z;\cdot,\cdot)\}_{z}$ verifies the conditions $(1)^{(n+1)}-(4)^{(n+1)}$.

Now we have to check the same properties for the kernel ${K}^{(n+1)}(z;\cdot,\cdot)$. To this end we introduce
\begin{equation}\label{B-K-int}
B(z;q,k|y>\kappa):= \int_{\Gamma_y :=\{q': \omega(q') = y\}}  K^{(n)}(z;q,q') \ K^{(1)}(z;q',q) \ d\nu_{y}(q') \ ,
\end{equation}
with integration over the Gelfand-Leray measure $\nu_{y}(\cdot)$ on the surface $\Gamma_y$, see \cite{AVG-Z}. Notice
that by (\ref{B-K-int}) one gets:
\begin{equation}\label{B-estim-1}
|B(z;q,k|y)| < \|{K}^{(n)}(z;\cdot,\cdot)\|_{{\mathcal{K}}_1}  \|{K}^{(1)}(z;\cdot,\cdot)\|_{{\mathcal{K}}_1} \,
h(q) h(k) \int_{\Gamma_y} h(q')^2 \ d\nu_{y}(q') \ .
\end{equation}
Thus, $B \in {\mathcal{K}}_1$ for any $z\in \overline{D_{1,\eta} \backslash I}$ and $y>\kappa$ with the
norm-estimate:
\begin{equation}\label{B-estim-2}
\|B(z;\cdot,\cdot|y)\|_{{\mathcal{K}}_1} < \|{K}^{(n)}(z;\cdot,\cdot)\|_{{\mathcal{K}}_1}
\|{K}^{(1)}(z;\cdot,\cdot)\|_{{\mathcal{K}}_1} \int_{\Gamma_y} h(q')^2 \ d\nu_{y}(q') \ ,
\end{equation}
and
\begin{equation}\label{B-estim-3}
\int_{0}^{\infty}  \|B(z;\cdot,\cdot|y)\|_{{\mathcal{K}}_1} \, dy <
\|{K}^{(n)}(z;\cdot,\cdot)\|_{{\mathcal{K}}_1} \|{K}^{(1)}(z;\cdot,\cdot)\|_{{\mathcal{K}}_1}
\int_{{\mathbb{R}}^d} h(q')^2 \ dq' \ .
\end{equation}

Now we can prove the following estimates of $y$-derivative of $B$:
\begin{equation}\label{B-estim-4}
\|\partial_{y} B(z;\cdot,\cdot|y)\|_{{\mathcal{K}}_1} <
C(y) \|{K}^{(n)}(z;\cdot,\cdot)\|_{{\mathcal{K}}_1} \|{K}^{(1)}(z;\cdot,\cdot)\|_{{\mathcal{K}}_1}
\end{equation}
where the asymptotic of the function $C(y)$ for $y \searrow \kappa$ is
\begin{equation}\label{asympt}
C(y) = (y - \kappa)^{(d/2 - 2)} + \mathcal{O} ((y - \kappa)^{(d/2 - 2)-\epsilon}) \ , \  \ \epsilon>0 \ ,
\end{equation}
whereas out of this $\kappa$-vicinity the function $C(y)$ is bounded.
%%%%%%%%%%%%%%%%%%%%%%%%%%%%%%%%%%%%%%%%% Lemma %%%%%%%%%%%%%%%%%%%%%%%%%%%%%%%%%%%%%%%%%%%%%%%%%%%%%%%%
\begin{lemma}\label{LammaApp-1}
Let $f(q)$ be a smooth function on $\mathbb{R}^d$. Let us define the function
\begin{equation}\label{Int}
\mathcal{I}(y):= \int_{\Gamma_y}  f(q) \ d\nu_{y}(q) \ .
\end{equation}
Then we have
\begin{equation}\label{Int-estim}
|\partial_{y} \mathcal{I}(y)| < d^2 \  \int_{\Gamma_y} \left\{\frac{|(\nabla f)(q)|}{|(\nabla \omega)(q)|} +
|f(q)|\frac{|(\Delta \omega)(q)|}{|(\nabla \omega)(q)|^2}\right\} \ d\nu_{y}(q) \ .
\end{equation}
\end{lemma}
%%%%%%%%%%%%%%%%%%%%%%%%%%%%%%%%%%%%%%%%%%%%%%%%%%%%%%%%%%%%%%%%%%%%%%%%%%%%%%%%%%%%%%%%%%%%%%%%%%%%%%%
{\it Proof.}  Notice that using \textit{differential forms} \cite{GS} one can rewrite (\ref{Int}) as
\begin{equation}\label{diff-form-1}
\mathcal{I}(y):= \int_{\Gamma_y}  \Omega_{0}(f) \ ,
\end{equation}
where $\Omega_{0}(f) = f \, \widetilde{\Omega}$ and $\widetilde{\Omega}$ is a special (Gelfand-Leray form),
which has the following \textit{local} coordinate representation:
\begin{equation}\label{diff-form-2}
\widetilde{\Omega} = \frac{1}{\partial \omega/\partial q_{1}} \ dq_2 \ldots dq_d \ .
\end{equation}
Without lost of generality we can suppose that $\partial \omega/\partial q_{1} \neq O$ and
$|\partial \omega/\partial q_{1}| \geq |\partial \omega/\partial q_{j}| \ , \ j = 2, \ldots , d $. By
(\ref{diff-form-1}) and (\ref{diff-form-2}) one gets \cite{GS}:
\begin{equation}\label{deriv-diff-form}
\partial_{y}\mathcal{I}(y) = \int_{\Gamma_y}  \Omega_{1}(f) \ , \ \ \ \
\Omega_{1}(f) = \left\{\partial_{q_1}f \ \frac{1}{\partial \omega/\partial q_{1}} +
f \ \partial_{q_1} \frac{1}{\partial \omega/\partial q_{1}}\right\} \widetilde{\Omega} \ .
\end{equation}
Since the convexity of $\omega$ implies $d |\partial \omega/\partial q_{1}| \geq |\nabla \omega|$ and
$|\partial^2 \omega/\partial q_{1}^{2}| < |\Delta \omega|$, we obtain the estimate
(\ref{Int-estim}). \hfill $\square$
%%%%%%%%%%%%%%%%%%%%%%%%%%%%%%%%%%%%%%%%% Corollary %%%%%%%%%%%%%%%%%%%%%%%%%%%%%%%%%%%%%%%%%%%%%%%%%%%%%%%%%%%%
\begin{corollary}\label{coroll-App-1}
The estimate (\ref{Int-estim}) implies (\ref{B-estim-4}) with
\begin{eqnarray}\label{C(y)}
C(y)&=& d^2 \  \int_{\Gamma_y} \left\{\frac{2 h(q)^2}{|(\nabla \omega)(q)|} \ + \
h(q)^2 \,\frac{|(\Delta \omega)(q)|}{|(\nabla \omega)(q)|^2}\right\} \ d\nu_{y}(q) = \\
&=& d^2 \ \partial_y \int_{\{\kappa < \omega(q) <y\}} h(q)^2 \left\{\frac{2}{|(\nabla \omega)(q)|} \ + \
\frac{|(\Delta \omega)(q)|}{|(\nabla \omega)(q)|^2}\right\} \ dq \ . \nonumber
\end{eqnarray}
Moreover, since for $y\searrow \kappa$ one gets in domain of integration: $|(\nabla \omega)(q)|\sim \sqrt{y - \kappa}$
and $ dq \sim |y - \kappa|^{d/2 - 1} dy$, the right-hand side of the last identity in (\ref{C(y)}) has asymptotics
$(y - \kappa)^{(d/2 - 2)}$ that proves (\ref{asympt}). On the other hand, for
$y > \kappa + \epsilon , \ \epsilon \geq \epsilon_0 > 0$ the same expression is bounded.
\end{corollary}
%%%%%%%%%%%%%%%%%%%%%%%%%%%%%%%%%%%%%%%%%%%%%%%%%%%%%%%%%%%%%%%%%%%%%%%%%%%%%%%%%%%%%%%%%%%%%%%%%%%%%%%%%%%

By virtue of (\ref{B-estim-4}) and (\ref{asympt}) we get for $d=3$ the $B(z;\cdot,\cdot|y)$ verifies the
H\"{o}lder condition with exponent $1/2$:
\begin{equation}\label{Hold-B-d3}
\|B(z;\cdot,\cdot|y_1) - B(z;\cdot,\cdot|y_2)\|_{{\mathcal{K}}_1} < \widehat{C}_{3} \ |y_1 - y_2|^{1/2} \ ,
\end{equation}
whereas for $d>3$ we obtain:
\begin{equation}\label{Hold-B-d>3}
\|B(z;\cdot,\cdot|y_1) - B(z;\cdot,\cdot|y_2)\|_{{\mathcal{K}}_1} < \widehat{C}_{>3} \ |y_1 - y_2| \ .
\end{equation}

Finally, $B(z;\cdot,\cdot|y)$ is continuous in the closure $\overline{D_{1,\eta} \backslash I}$, as a function of $z$,
and by (\ref{B-K-int}) we have for it on the cut $I$ the values:
\begin{equation}\label{B-K-int-cut}
B^{\pm}(x;q,k|y):= \int_{\Gamma_y}  K^{(n),\pm}(x;q,q') \ K^{(1),\pm}(x;q',q) \ d\nu_{y}(q') \ ,
\end{equation}
which satisfy the above estimates (\ref{B-estim-2})-(\ref{B-estim-4}) and (\ref{Hold-B-d3}) or (\ref{Hold-B-d>3}),
for $z=x$.

Moreover, we find behaviour of (\ref{B-K-int-cut}) as a function of $x$:
\begin{equation}\label{funct-x}
\|B(x_1;\cdot,\cdot|y) - B(x_2;\cdot,\cdot|y)\|_{{\mathcal{K}}_1} < \tilde{C}_{d} \ |x_1 - x_2|^{\gamma_d} \ ,
\end{equation}
where $\tilde{C}_{d}= L_n \, C_1 + L_1 \, C_n$ (see (\ref{estim})) and $\gamma_{d=3} =1/2$, $\gamma_{d>3} = 1$ .

Summarizing we conclude that for $j=0$ we obtain the corresponding representation of the kernel (\ref{K-bis})
in the form:
\begin{equation}\label{repr-0}
K^{(n+1),0}(z;q,k) = \int_{\kappa}^{\infty} \, \frac{B(z;q,k|y)}{e_0 + y - z}\ dy  \  .
\end{equation}
Now to establish desired properties, $(1)^{(n+1)}-(4)^{(n+1)}$, of this kernel on the basis of the $B(z;q,k|y)$ properties,
we consider a more general integral:
\begin{equation}\label{repr-1}
\widetilde{K}^{(n+1),0}(z_1, z_2;q,k) = \int_{\kappa}^{\infty} \, \frac{B(z_1;q,k|y)}{e_0 + y - z_2}\ dy  \  .
\end{equation}
%%%%%%%%%%%%%%%%%%%%%%%%%%%%%%%%%%%%%%%%% Lemma %%%%%%%%%%%%%%%%%%%%%%%%%%%%%%%%%%%%%%%%%%%%%%%%%%%%%%%%
\begin{lemma}\label{LammaApp-2} (Privalov's lemma for vector-valued functions)\\
Let $\{f(z,y)\}_{z,y} \subset \mathcal{B}$ for $z\in \mathcal{D}_{0}\setminus I$ with a cut $I$ and
$y\in (\kappa, \infty)$ be family of vector-valued functions in a Banach space $\mathcal{B}$.
Assume that they verify the following conditions:\\
{\rm{(a)}} $f(z,y)$ is $\mathcal{B}$-analytic in $\mathcal{D}_0\setminus I$ and $\mathcal{B}$-continuous on the
closure $\overline{\mathcal{D}_0\setminus I}$ for any fixed $y\in (\kappa, \infty)$.\\
{\rm{(b)}} The limit values $f^{\pm}(x,y)$ on the cut $I$ verify the H\"{o}lder condition:
\begin{equation}\label{Hold-Ban-1}
\|f^{\pm}(x_1,y) - f^{\pm}(x_2,y)\|_{\mathcal{B}} < C_{1} |x_1 - x_2|^{1/2} \ , \  x_{1,2} \in I \ .
\end{equation}
{\rm{(c)}} For any $z\in \overline{\mathcal{D}_0\setminus I}$ one has
\begin{equation}\label{Hold-Ban-2}
\|f(z,y_1) - f(z,y_2)\|_{\mathcal{B}} < C_{2}|y_1 - y_2|^{1/2} \ , \  y_{1,2} \in (\kappa, \infty) \ .
\end{equation}
{\rm{(d)}} For any $z\in \overline{\mathcal{D}_0\setminus I}$ the integral
\begin{equation}\label{summable}
\int_{\kappa}^{\infty} \, \|f(z,y)\|_{\mathcal{B}} \ dy < R \ .
\end{equation}
{\rm{(f)}} Uniform boundedness and limits at infinity:
\begin{equation}\label{Bound}
\sup_{\{z\in \overline{\mathcal{D}_0\setminus I}, y \in (\kappa, \infty)\}} \|f(z,y)\|_{\mathcal{B}} < M
\  \ {\rm{and}} \ \ \lim_{z\rightarrow\infty} \|f(z,y)\|_{\mathcal{B}} =
\lim_{y\rightarrow\infty} \|f(z,y)\|_{\mathcal{B}} = 0 \ .
\end{equation}
Let us define the (Bochner) integral
\begin{equation}\label{Boch-Int}
F(z_1,z_2):= \int_{\kappa}^{\infty} \, \frac{f(z_1,y)}{e_0 + y - z_2} \ dy \ ,
\end{equation}
where $e_0$ is defined by the cut $I$ (\ref{cut-interval}).
Then the function  (\ref{Boch-Int}) has the following properties:\\
{\rm{(1)}} It is analytic in $\mathcal{D}:=\{\mathcal{D}_0\setminus I\} \times \{\mathcal{D}_0\setminus I\}$ in
two variables $z_1,z_2$ .
{\rm{(2)}} It is uniformly bounded and continuous in the closure $\overline{\mathcal{D}}$:
\begin{equation*}
\|F(z_1,z_2)\|_{\mathcal{B}} < A_1 \, M \ .
\end{equation*}
{\rm{(3)}} The limit values $F^{\pm}(x_1,x_2)$ on the cut $I$ satisfy the H\"{o}lder conditions:
\begin{equation*}
\|F^{\pm}(x_1 + \delta_1, x_2 + \delta_2) - F^{\pm}(x_1, x_2)\|_{\mathcal{B}} <
\widehat{C} (|\delta_1|^{1/2} + |\delta_2|^{1/2}) \ ,
\end{equation*}
where
\begin{equation*}
F^{\pm}(x_1,x_2) := \lim_{\varepsilon_{1,2}\rightarrow +0} F(x_1 \pm i \varepsilon_1 ,x_2 \pm i \varepsilon_2) \ ,
\end{equation*}
and $ \ \widehat{C} = A_2 (C_{1} + C_{2})$ .
\end{lemma}
%%%%%%%%%%%%%%%%%%%%%%%%%%%%%%%%%%%%%%%%%%%%%%%%%%%%%%%%%%%%%%%%%%%%%%%%%%%%%%%%%%%%%%%%%%%%%%%%%%%%%%%
\textit{Proof.} Follows through verbatim of the standard demonstration for complex-valued functions, see
e.g. \cite{Pr}. \hfill $\square$
%%%%%%%%%%%%%%%%%%%%%%%%%%%%%%%%%%%%%%%%%%%%%%%% Proof Proposition 3.1 %%%%%%%%%%%%%%%%%%%%%%%%%%%%%%%%%

Applying Lemma \ref{LammaApp-2} to our case of $\mathcal{B} =\mathcal{K}_1$ shows that limit values
$\widetilde{K}^{(n+1),0,\pm}(x_1, x_2; q,k)$ of integral (\ref{repr-1}) verify the H\"{o}lder condition for
variables $x_1, x_2$. Since
\begin{equation*}
{K}^{(n+1),0}(z; q,k)=\widetilde{K}^{(n+1),0,\pm}(z, z; q,k) \ ,
\end{equation*}
we obtain that the function ${K}^{(n+1),0,\pm}(x; q,k)$ also verifies the H\"{o}lder condition for the
variable $x$. Moreover, from the estimates that contain the factor
$\max_{\{z\in\overline{D_{1,\eta} \backslash I}\}} \|{K}^{(1)}(z; \cdot,\cdot)\|_{\mathcal{K}_1} = \alpha$
yield:
\begin{equation*}
\|{K}^{(n+1),0}(z; \cdot,\cdot)\|_{\mathcal{K}_1} < const \ \alpha \|{K}^{(n)}(z; \cdot,\cdot)\|_{\mathcal{K}_1}
\end{equation*}
and
\begin{equation*}
\|{K}^{(n+1),0,\pm}(x_1; \cdot,\cdot) - {K}^{(n+1),0,\pm}(x_2; \cdot,\cdot)\|_{\mathcal{K}_1} < const \ \alpha \ C_n \ ,
\end{equation*}
see (\ref{estim}).

Similarly one checks these estimates for the family $\{{K}^{(n+1),1}(z; \cdot,\cdot)\}_{z}$. Therefore, the whole
family $\{{K}^{(n+1)}(z; \cdot,\cdot)\}_{z}$ verifies the recurrent estimates (\ref{estim}), that finishes the proof
of Proposition \ref{P3.1}. \hfill $\square$

\newpage
%%%%%%%%%%%%%%%%%%%%%%%%%%%%%%%%%%%%%%%%%%%%% Biblio %%%%%%%%%%%%%%%%%%%%%%%%%%%%%%%%%%%%%%%%%%%%%%%%%%%%%%

\newpage

\section*{Figures Captions}

\begin{itemize}
 \item
Figure 1a: The graph of the l.h.s.\ of (\ref{e3.2}) : case of two roots
\item
Figure 1b: The graph of the l.h.s.\ of (\ref{e3.2}) : case of one root
\item
Figure 2: The graph of the l.h.s.\ of (\ref{e3.10})
\item
Figure 3:
Solving Eq. (\ref{e3.26})
\end{itemize}

\newpage
\thispagestyle{empty}
\includegraphics{figamrz-1a.eps}

\newpage
\thispagestyle{empty}
\includegraphics{figamrz-1b.eps}

\newpage
\thispagestyle{empty}
\includegraphics{figamrz-2.eps}

\newpage
\thispagestyle{empty}
\includegraphics{figamrz-3.eps}

\newpage
\thispagestyle{empty}

%%%%%%%%%%%%%%%%%%%%%%%%%%%%%%%%%%%%%%%%%%%%%%%%%%%%%%%%%%%%%%%%%%%%%%%%%%%%%%%%%%%%%%%%%%%%%%%%%%%%%
\end{document}